\documentclass[aps,prd,preprint,nofootinbib,superscriptaddress]{revtex4}
\usepackage{color}
\usepackage{graphicx}
\usepackage{amsmath}
\usepackage{epsfig}

\def\lsim{\:\raisebox{-0.9ex}{$\stackrel{\textstyle<}{\sim}$}\:}
\def\gsim{\:\raisebox{-0.9ex}{$\stackrel{\textstyle>}{\sim}$}\:}
\def\half{\frac{1}{2}}

\def\cm{\,{\rm cm}}

\def\km{\,{\rm km}}

\def\s{\,{\rm s}}

\def\kg{\,{\rm kg}}

\def\kmps{\km\s^{-1}}

\def\pb{\,{\rm pb}}

\def\erf{{\rm erf}}

\def\sigmachiAzero{\sigma_{\chi A}(0)}
\def\sigmachiASIzero{\sigma_{\chi A}^{\rm SI}(0)}

\def\sigmachinSI{\sigma_{\chi n}^{\rm SI}}
\def\sigmachinSIUL{\sigma_{\chi n,90}^{\rm SI}}

\def\GeV{\,{\rm GeV}}

\newcommand{\etal}{{\it et al.\ }}

\def\hydrogen1{$^1$H}
\def\carbon12{$^{12}$C}
\def\fluorine19{$^{19}$F}

\def\c2h2f4{C$_{2}$H$_{2}$F$_{4}$}  
\def\cthreef8{C$_{3}$F$_{8}$}
\def\cf3I{CF$_{3}$I}

\def\Eth{E_{\rm th}}
\def\ER{E_{\rm R}}

\def\ERmaxi{E_{\rm R, max}^{(i)}}
\def\ERth{E_{\rm R, th}}
\def\ERthi{E_{\rm R, th}^{(i)}}

\def\Rc{R_{\rm c}}
\def\Ec{E_{\rm c}}
\def\Tth{T_{\rm th}}
\def\Edep{E_{\rm dep}}
\def\Edep2Rc{E_{\rm dep}^{2\Rc}}

\def\etaT{\eta_{\rm T}}
\def\mathcalR{\mathcal{R}}
\def\mathcalE{\mathcal{E}}
\def\mchi{m_{\chi}}
\def\mchilowest{m_{\chi, {\rm lowest}}} 
\def\mA{m_A}
\def\mAi{m_{A_i}}
\def\muchiA{\mu_{\chi A}}

\def\vmin{v_{\rm min}}
\def\vesc{v_{\rm esc}}
\def\vE{v_{\rm E}}

\def\boldmathv{{\boldsymbol{v}}}
\def\boldmathvE{{\boldmathv}_{\mathrm E}}

\def\vdispsq{\langle v^2 \rangle}

\def\rhodm{\rho_{\rm\scriptscriptstyle DM}}

\begin{document}

\title{Probing low-mass WIMP candidates of dark matter with 
tetrafluoroethane superheated liquid detectors}

\author{Susnata Seth\footnote{Presently at Bose Institute, EN-80, 
Sector V, Bidhannagar, Kolkata 700091, India.}}
\email{susnata.seth@gmail.com}
\affiliation{Astroparticle Physics \& Cosmology Division,
Saha Institute of Nuclear Physics, Kolkata 700064, India}
\author{Sunita Sahoo}
\email{sunita.sahoo@saha.ac.in}
\affiliation{Astroparticle Physics \& Cosmology Division,
Saha Institute of Nuclear Physics, Kolkata 700064, India}
\affiliation{Homi Bhabha National Institute, Training
School Complex, Anushakti Nagar, Mumbai 400094, India}
\author{Pijushpani Bhattacharjee}
\email{pijush.bhattacharjee@saha.ac.in}
\affiliation{Astroparticle Physics \& Cosmology Division,
Saha Institute of Nuclear Physics, Kolkata 700064, India}
\author{Mala Das}
\email{mala.das@saha.ac.in}
\affiliation{Astroparticle Physics \& Cosmology Division,
Saha Institute of Nuclear Physics, Kolkata 700064, India}
\affiliation{Homi Bhabha National Institute, Training
School Complex, Anushakti Nagar, Mumbai 400094, India}
\begin{abstract}
\noindent 
Probing low mass (sub-GeV -- few GeV) Weakly Interacting Massive 
Particle (WIMP) candidates of dark matter through WIMP-induced nuclear 
recoils in direct detection experiments requires use of detector 
materials consisting of low mass target nuclei and low 
threshold energy. Here we explore the potential of superheated liquid 
detectors (SLD) with a hydrogen containing liquid, namely, 
tetrafluoroethane (\c2h2f4) (b.p.~$-26.3^\circ\,$C), as the target 
material for probing low mass WIMPs. It is found that few-keV level 
recoil energy thresholds possible for bubble nucleation by WIMP-induced 
\carbon12 and \fluorine19 recoils in \c2h2f4 SLDs operated at 
atmospheric pressure and 
gamma-ray insensitive temperatures of $T\, \lsim\, 35^\circ\,$C have  
the potential to allow WIMPs in the few-GeV mass range to be probed at a 
WIMP-nucleon spin-independent cross section sensitivity levels (90\% 
C.L.) better than $4.6\times 10^{-5}\pb$ at WIMP masses down to $\sim$ 4 
GeV with a total exposure of $\sim$ 1000 kg.day, provided that the 
``thermodynamic efficiency" $\etaT$ that 
determines the bubble nucleation thresholds for the recoiling nuclei in 
\c2h2f4 is $\sim$ 50\% or higher. Sensitivity 
to sub-GeV WIMP masses generally requires the detector to be 
sensitive to the WIMP-induced \hydrogen1 recoils, which in turn requires 
the detector to be operated at temperatures $T\gsim 50^\circ\,$C and 
$\etaT$ close to 100\%. At such relatively high temperatures (at 
atmospheric pressure), however, the detector would be sensitive to 
background gamma rays.   

\end{abstract}
\maketitle

\section{Introduction}\label{sec:intro}
Weakly Interacting Massive Particles (WIMPs)~\cite{wimp1,wimp2,wimp3}  
predicted in many theories beyond the Standard Model of particle 
physics, with masses of a few GeV to a few hundred TeV\footnote{We 
use units with $c=1$ throughout this paper.} are 
one of the major candidates for the constituents 
of the Dark Matter (DM), an unknown form of non-luminous matter that 
constitutes about 85\% of the total gravitating mass and about 27\% of 
the total mass-energy budget of the Universe; 
see, e.g., Ref.~\cite{wimp-dm-review} for a recent review. 
Following the early suggestion~\cite{Goodman-Witten} that nuclear recoil 
events due to elastic scattering of the Galactic WIMPs off nuclei of 
suitably chosen detector materials may be detectable, 
a large number of experiments worldwide have been engaged for the past 
three decades or so in efforts to detect the WIMPs employing various 
detection techniques. The kinetic energy of a recoiling nucleus due to 
WIMP-nucleus elastic scattering, which can be anywhere in the range of a 
few keV to few hundreds of keV depending on the WIMP 
and target nucleus masses, would be dissipated in the detector 
medium providing signals in the form of bolometric heat, lattice 
vibration (phonon), ionization, scintillation light, and so on, 
depending on the detector 
medium~\cite{wimp-dd-signal-Gaitskell-rev,wimp-dd-rev-Undagoitia-etal}. 
The DAMA/LIBRA 
experiment~\cite{dama-2008,dama-libra-2010,dama-libra-2013} has 
been consistently reporting, for about a decade now, a 
statistically significant detection of an annual modulation 
signal in their event rate, which they attribute to WIMPs, the 
annual modulation being attributed to Earth's motion around the 
Sun~\cite{annual-mod1,annual-mod2}. However, the DAMA/LIBRA results are 
difficult to reconcile with the null results from a  
number of other experiments which have set rather stringent upper limits 
on the WIMP-nucleon interaction 
strength~\cite{cresst-II-2016,pico-60-C3F8-PRL2017,lux-2017,xenon-1t-2017,pandaX-II-2017,supercdms-lite-2018}. 

Most of the currently running experiments are designed to 
be optimally sensitive to WIMPs of mass $\gsim 10\GeV$. In view of the 
null results from these experiments, recently there has been much 
interest in experiments designed to be specifically sensitive to 
relatively lower mass ($ < 10\GeV$) WIMPs; see, for example, 
Refs.~\cite{cresst-II-2016,supercdms-lite-2018,cdex-10-2018,DarkSide-50-2018,NEWS-G-2018}. 
Sensitivity to low mass WIMPs generally require low (sub-keV) recoil 
energy threshold and detector materials containing low mass nuclei. 

In this paper we study the possibility of probing low mass 
(sub-GeV -- few GeV) WIMPs with a detector material containing hydrogen, 
the lowest-mass target nucleus possible. Specifically, we consider the 
superheated liquid tetrafluoroethane (\c2h2f4), a low-cost, commercially 
available, environment friendly (chlorine free) refrigerant 
liquid (b.p.~-26.3$^{\rm o}$C) as the target detector material. 

Superheated liquid based detectors with liquids such as 
C$_{3}$F$_{8}$, CF$_{3}$I, C$_{4}$F$_{10}$, C$_{2}$ClF$_{5}$, and so 
on have been extensively used for WIMP direct detection 
experiments~\cite{pico-60-C3F8-PRL2017,picasso-2011,simple-2014,picasso-2017,moscab-2017}. 
The superheated liquid state being a metastable state of the 
liquid~\cite{Skirpov}, the energy deposited by a recoiling nucleus 
arising from WIMP-nucleus scattering in the liquid can induce a 
phase transition from the superheated liquid state to vapor state if 
the deposited energy exceeds a certain critical energy that depends on 
the temperature and pressure of the liquid. The acoustic pulse  
generated during such a phase transition provides the signal which
can be detected by acoustic sensors. The phase transition from the 
superheated liquid state to the vapor state occurs through nucleation of 
vapor bubbles of certain critical size. Since bubble 
nucleation can occur only if the energy deposition is above a certain 
critical amount, this makes such detectors act as 
threshold detectors, with the threshold energy controlled by the 
temperature and pressure of the liquid. A major advantage of superheated 
liquid 
detectors is their operability at room temperatures as opposed to 
cryogenic temperatures required for most other kinds of 
WIMP search detectors currently under operation. Moreover, by 
controlling the threshold energy of the detector with 
judicious choice of the operating temperature and pressure, the detector 
can be made 
insensitive to certain kinds of particles, for example, beta particles and 
gamma-rays, which constitute the main sources of background for 
most WIMP search experiments. 

Gamma ray and neutron sensitivities of superheated liquid 
detectors with \c2h2f4 as the active liquid  
have been studied earlier in various contexts including neutron 
dosimetry and spectrometry; see, e.g., 
~\cite{bubble-nucl-dErrico-2001,c2h2f4-Mala-Pramana-2010,c2h2f4-PrasannaMondal-2014,c2h2f4-SunitaSahoo-2019}. 
Recently, the PICO collaboration has started 
exploring the possibility of using \c2h2f4 for low-mass WIMP 
detection~\cite{PICO-int-Plante-Zacek-2016}. Work has also been done   
studying bubble nucleation due to proton recoils in \c2h2f4 using 22.8 
keV neutrons from a $^{124}$SbBe source~\cite{PICO-thesis-Tardif-2018}. 
Our aim in this 
paper is to present a theoretical study of the response of superheated 
liquid \c2h2f4 detectors to low-mass WIMPs within the general context of 
Seitz's~\cite{Seitz} phenomenological theory of bubble nucleation in 
superheated liquids.  
In particular, we theoretically study the behavior of the bubble 
nucleation threshold energies of the WIMP-induced 
recoiling hydrogen (\hydrogen1), carbon (\carbon12) and fluorine 
(\fluorine19) nuclei in \c2h2f4 at various temperatures and the corresponding 
lowest WIMP mass that can be probed with \c2h2f4 detectors. 

Below, in section \ref{sec:detector-principle} we briefly review the 
basic working principle of superheated liquid detectors (SLD) and 
discuss the method we follow to calculate the bubble nucleation 
threshold energies of various particles moving through the liquid. 
In section \ref{sec:response-to-wimps} we discuss the response of SLD to 
spin-independent elastic scattering of the WIMPs constituting the DM 
halo of our Galaxy, focusing on the lowest WIMP mass to 
which the SLD can be sensitive. Section 
\ref{sec:results} presents our results for the bubble nucleation 
threshold energies of recoiling hydrogen (\hydrogen1), carbon 
(\carbon12) and fluorine (\fluorine19) nuclei in superheated liquid 
\c2h2f4 and the corresponding lowest WIMP mass that can be probed        
with \c2h2f4 SLD as a function of temperature. Finally,  
section~\ref{sec:summary} summarizes our main results and conclusions.    

\section{Superheated Liquid Detectors: Basic 
Principles}\label{sec:detector-principle}
A Superheated Liquid Detector (SLD) works on the basic principle that 
localized energy deposition during the passage of an energetic 
particle through the liquid can cause a phase transition from 
the liquid state to the vapor phase. 
According to Seitz's phenomenological ``heat spike" theory~\cite{Seitz}, 
the phase transition occurs through nucleation of vapor bubbles of radii 
larger than a critical radius ($\Rc$) due to localized deposition 
of energy by the particle within the superheated liquid. The 
bubbles of radii smaller than $\Rc$ collapse back to the liquid 
state while those with radii larger than $\Rc$ expand and grow to 
visible size through evaporation of the liquid. The expansion of the 
vapor bubble is accompanied by production of an acoustic pulse which 
acts as the signal carrying information about the 
energy deposited in the liquid due to the passage of the particle. 

At a given temperature and pressure, the critical radius ($\Rc$) is 
given by~\cite{Seitz} 
\begin{equation}
\Rc=\frac{2\sigma(T)}{(P_{v}-P_{l})}, 
\label{eq:Rc_def}
\end{equation}
where $\sigma(T)$ is the liquid-vapor interfacial tension at 
temperature $T$, $P_{v}(T)$  is 
the vapor pressure and $P_{l}(T)$  is 
the pressure of the liquid. To form a bubble of critical radius 
the particle must have an energy\footnote{Throughout this paper, we 
shall be concerned with particles of non-relativistic speeds, and hence 
by energy of a particle we shall mean its non-relativistic kinetic 
energy.}, $E$, equal to or greater than a certain 
threshold energy, $\Eth$ such that the energy deposited by the particle, 
$\Edep2Rc$, over a path segment of length $2\Rc$ (the ``critical 
diameter") along the particle's track in the liquid
satisfies the condition 
\begin{equation}
\Edep2Rc (E=\Eth)\equiv\int_{0}^{2\Rc}
\left(\frac{dE}{dx}\right) dx = \Ec/\etaT\,,
\label{eq:E_dep_condition}
\end{equation}
where $\frac{dE}{dx}$ is the stopping power of the liquid for the
particle under consideration, $\etaT \leq 1$ is the ``thermodynamic
efficiency"~\cite{Apfel-etal-1985} and $\Ec$ is the minimum
(``critical") energy required for bubble nucleation, which is given by
\cite{Seitz,pico-60-CF3I-PRD2016}
\begin{equation}
\Ec = 4\pi \Rc^2\left(\sigma-T\frac{\partial \sigma}{\partial 
T}\right)+\frac{4\pi}{3}\Rc^3\rho_{v}\left(h_{v}-h_{l}\right)
 -\frac{4\pi}{3}\Rc^3\left(P_{v}-P_{l}\right)\,,
\label{eq:Ec_def}
\end{equation}
where $\rho_{v}(T)$ is the vapor density, and $h_{v}(T)$, $h_{l}(T)$ 
are the specific enthalpies of the vapor bubble and liquid, 
respectively. Equation (\ref{eq:E_dep_condition}) also serves to
define the thermodynamic efficiency $\etaT$ as a measure of the
fraction of the energy deposited by the particle within $2\Rc$ that goes
into nucleation of a bubble of critical radius
$\Rc$, and is a characteristic of the superheated liquid under
consideration. 

If at any given temperature and pressure the mean range\footnote{The
range is defined as the average distance over which the particle loses
all its kinetic energy and comes to a stop in the liquid.}, $R$, of the
particle at energy $E=\Ec/\etaT$ satisfies $R(E=\Ec/\etaT) \leq 2\Rc$,
then we have $\Eth = \Ec/\etaT$. On the other hand, if $R(\Ec/\etaT) >
2\Rc$, then $\Eth$ will be larger than $\Ec/\etaT$ and is determined by
equation (\ref{eq:E_dep_condition}). Also, since at
a given pressure the critical energy $\Ec$ decreases with increasing
temperature (see Table \ref{table:Ec-Rc-Leff-Range} below), the energy
threshold $\Eth$ translates to a
temperature threshold, $\Tth$, for bubble nucleation, with lower $\Eth$
corresponding to higher $\Tth$ and vice versa.

Clearly, the threshold energy for bubble nucleation depends on the value
of the thermodynamic efficiency $\etaT$, which is {\it a priori}
unknown, and can only be determined through experiment. The Seitz
theory~\cite{Seitz} generally assumes $\etaT=1$, in which case $\Eth\geq
\Ec$. For this reason, the energy $\Ec$ is often referred to as the
``Seitz threshold". Recent results from the PICO
experiment~\cite{pico-60-CF3I-PRD2016,pico-60-C3F8-PRD2019} seem to
suggest that the actual bubble nucleation threshold energy $\Eth$ of
different nuclei can in fact be significantly larger than the Seitz 
threshold $\Ec$ (as defined by equation (\ref{eq:Ec_def})), with the 
lighter nuclei generally having larger values of the ratio $\Eth/\Ec$. 
From Figure 3 of Ref.~\cite{pico-60-C3F8-PRD2019} we see that, in terms 
of our parameter $\etaT$, the $1\sigma$ bands of the experimental values 
of $\Eth/\Ec$ for bubble nucleation thresholds of \carbon12 and 
\fluorine19 recoils in \cthreef8 would correspond to values of $\etaT$ 
($=\Ec/\Eth$ in these cases --- see the discussions in the previous 
paragraph) roughly in the range $\sim$ 0.3 -- 0.9. 
Taking cues from these results, we may expect similar values of the 
parameter $\etaT$ to be applicable in the case of \c2h2f4 too, at least 
for \carbon12 and \fluorine19, since both these nuclei behave 
roughly similarly in the two liquids as far as their energy deposits 
(determined by the ratio of their range $R$ to the critical radius $\Rc$ 
in the liquid under consideration) are concerned. Based on the above 
considerations, in absence of 
any direct experimental results on the threshold energies of individual 
nuclei in the case of \c2h2f4, in this paper we shall do the 
calculations and show the results for the event rates and WIMP-mass 
sensitivities for \c2h2f4 for two different values of $\etaT$, namely, 
50\% and 100\%, for illustration.  

The threshold energy $\Eth$ marks the {\it onset} of bubble nucleation, 
with no bubble nucleation occurring below this energy. However, in 
reality, the process of bubble nucleation being a probabilistic one, 
the efficiency of bubble 
nucleation may not be 100\% at the energy $\Eth$ itself, but may reach 
full efficiency only at a somewhat higher value of energy, as seen in 
experiments~\cite{pico-60-CF3I-PRD2016,pico-60-C3F8-PRD2019}. 
The nucleation efficiency curves, i.e., the bubble nucleation 
efficiency as a function of the ion's recoil energy, of individual ions 
in \c2h2f4 are not known. However, the PICO 
experiment~\cite{pico-60-C3F8-PRD2019} has determined, albeit with large 
uncertainties, the bubble nucleation efficiency curves of \carbon12 and 
\fluorine19 recoils in \cthreef8. To make progress, based on the 
discussions in the previous paragraph, below we shall use these  
efficiency curves for \carbon12 and \fluorine19 
in \cthreef8, appropriately scaled to their respective threshold 
energies in \c2h2f4, for our calculations of the expected event rates 
and WIMP mass sensitivities of \c2h2f4 due to \carbon12 and \fluorine19 
recoils. For \hydrogen1 recoils, in absence of any direct experimental 
results on their efficiency curves, we shall assume 100\% bubble 
nucleation efficiency at threshold as in 
previous studies~\cite{PICO-thesis-Tardif-2018}. 

\section{Response of Superheated Liquid Detector to WIMPs}
\label{sec:response-to-wimps}
In a direct detection experiment for DM search, the detector looks for 
signatures of nuclear recoils produced by scattering of the WIMPs 
off the nuclei of the detector material. The recoil energy of 
the nucleus, $\ER$, due to WIMP-nucleus elastic scattering is given by  
\begin{equation}
\ER=\frac{\muchiA^2v^2}{\mA}(1-\cos\theta),
\label{eq:ER_WIMP}
\end{equation} 
where $v=|{\boldmathv}|$ is the speed of the WIMP relative to the 
target nucleus at 
rest on Earth, $\theta$ is the WIMP scattering angle in the 
WIMP-nucleus center-of-mass system, $\mchi$ and $\mA$ are the WIMP 
and target nucleus masses, respectively, and $\muchiA=\frac{\mchi 
\mA}{\mchi+\mA}$ is the WIMP-nucleus reduced mass. The minimum WIMP 
speed, $\vmin$, that 
can produce a recoil nucleus with energy $\ER$ is given by 
\begin{equation}
\vmin=\left( \frac{\ER \mA}{2\muchiA^2}\right)^{1/2}.
\label{eq:vmin}
\end{equation} 

The differential recoil rate, $d\mathcalR/d\ER$, i.e., the number 
of nuclear recoil events per unit time per unit 
detector mass per unit recoil energy can be written 
as~\cite{wimp2,Lewin-Smith-1996}  
\begin{equation}
\frac{d\mathcalR}{d\ER} = \frac{\sigmachiAzero\, 
\rhodm}{2\mchi\,\muchiA^2} 
F^2(q) 
\int_{\vmin(\ER)} d^3v \frac{f(\boldmathv,t)}{v}\,,
\label{eq:recoil-spect1}
\end{equation}
where $\sigmachiAzero$ is the `zero momentum transfer' WIMP-nucleus 
cross-section, $\rhodm\approx 0.3\GeV/\cm^3$ is the local mass
density of DM~\cite{Read-2014}, $f(\boldmathv,t)$ is the WIMP  
velocity distribution in the Earth's rest frame, the time dependence 
being due to Earth's revolution around the 
Sun~\cite{annual-mod1,annual-mod2}, and 
$F(q)$ (with $F(0)=1$) is the nuclear form factor with $q=(2\mA 
\ER)^{1/2}$ the momentum transfer from the WIMP to the nucleus.      

For the WIMPs' velocity distribution, we shall assume the standard halo 
model (SHM) in which the DM halo of the Galaxy is described by an 
isothermal sphere~\cite{Binney-Tremaine-2008} with an isotropic 
velocity distribution of the Maxwell-Boltzmann 
form in the Galactic rest frame truncated at the local Galactic escape 
speed $\vesc$ and Galilean boosted to the Earth's frame 
(see, e.g., ~\cite{annual-mod2,Lewin-Smith-1996}): 
\begin{equation}
f(\boldmathv,t)=\frac{1}{\kappa}\frac{1}{(\pi 
v_0^2)^{3/2}}\exp{\left\{-\frac{(\boldmathv + 
\boldmathvE)^2}{v_0^2}\right\}} \Theta (\vesc - |\boldmathv + 
\boldmathvE|)\,,
\label{eq:VDF}
\end{equation}
where $v_0=(\frac{2}{3}\vdispsq)^{1/2}\simeq 220\kmps$ is the 
characteristic (most probable) speed of 
the DM particles in the Galaxy, $\boldmathvE (t)$ is the 
velocity of the Earth with respect to the Galactic rest 
frame, 
\begin{equation}
\kappa=\erf\left(\frac{\vesc}{v_0}\right)
- \frac{2}{\sqrt{\pi}}\frac{\vesc}{v_0}\exp\left(\frac{-\vesc^2}{v_0^2}
\right)
\label{eq:kappa}
\end{equation}
is a normalization constant, and $\Theta(x)$ is the unit step function. 
The exact value of $\vesc$ is not 
known with certainty. Values in the range from 498 to 608 
$\kmps$ (90\% C.L.) are quoted in literature, with a median 
value of $\sim 540\kmps$ ~\cite{escape-speed}, which we shall use 
in this paper for all numerical estimates.  

In this paper we shall not consider the (small) annual modulation of the 
recoil rate (\ref{eq:recoil-spect1}) due to Earth's revolution around 
the Sun~\cite{annual-mod1,annual-mod2}, and consider only the annual 
average value of the recoil rate with the average value of 
$\vE=|\boldmathvE|\simeq 232\kmps$. 

With the WIMP speed distribution given by equation (\ref{eq:VDF}), the 
differential recoil rate (equation (\ref{eq:recoil-spect1})) for a 
detector 
consisting of nuclei of mass number $A$ and atomic number $Z$ can be 
written as~\cite{Lewin-Smith-1996}  

\begin{equation}
\begin{split}
\frac{d\mathcalR}{d\ER} & = \kappa^{-1}\, \frac{\mathcalR_0}{rE_0} 
F^2(q)\, 
\biggl[\frac{\sqrt{\pi}}{4}\frac{v_0}{\vE} 
\biggl\{\erf\biggl(\frac{\vmin+\vE}{v_0}\biggr)\\
& {} - \erf\biggl(\frac{\vmin-\vE}{v_0}\biggr)\biggr\}
  -\exp\biggl(\frac{-\vesc^2}{v_0^2}\biggr)\biggr]\,,
\label{eq:recoil-spect2}
\end{split}
\end{equation}
where $\mathcalR_0= \frac{2}{\sqrt{\pi}} \frac{N_0}{A} 
\frac{\rhodm}{\mchi}\sigmachiAzero \, v_0$, 
$N_0 = 6.022 \times 10^{26} \kg^{-1}$ is the Avogadro number,  
$r=\frac{4\mA\mchi}{(\mA+\mchi)^2}$, and $E_0=\half \mchi v_0^2$. 
Note, in equation (\ref{eq:recoil-spect2}), the $\ER$ 
dependence is contained in $\vmin$ [see equation (\ref{eq:vmin})] and 
the form factor $F(q)$. The latter can be taken to be of the form    
~\cite{Lewin-Smith-1996}
\begin{eqnarray}
F(q)=3\, e^{-(qs)^2/2}\,\,\,   
\frac{\sin(qr_n)-qr_n\cos(qr_n)}{(qr_n)^3}\,,
\label{eq:Form-factor}
\end{eqnarray}
where $r_n$ is the effective nuclear radius given by 
$r_n^2= c^2+\frac{7}{3}\pi^2a^2 - 5s^2$ with 
$c\simeq1.23A^{1/3}-0.60$ fm, and nuclear skin thickness parameters 
$a\simeq0.52$ fm and $s\simeq0.9$ fm.  

In this paper, we shall restrict ourselves to the case of coherent, 
spin-independent (SI) WIMP-nucleus interaction. In this case, assuming 
isospin independent WIMP-nucleon coupling, the zero momentum 
WIMP-nucleus cross section $\sigmachiAzero$ can be written in terms of 
the SI WIMP-nucleon cross section, $\sigmachinSI$, as   
\begin{equation}
\sigmachiAzero = \sigmachiASIzero = \sigmachinSI \,\, 
\frac{\left(1+\frac{\mchi}{m_n}\right)^2}{\left(1+\frac{\mchi}{\mA}\right)^2} 
\,\, A^2\,,
\label{eq:sigma0-sigmachin-reln}
\end{equation}
where $m_n$ is the nucleon (neutron or proton) mass. 

For a detector made of a compound target material consisting of 
different elements $i$ of mass numbers $A_i$ and nuclear masses 
$\mAi$, as is the case in this paper, the differential recoil 
rate (per unit mass of the compound target material) is given by 
\begin{equation}
\frac{d\mathcalR}{d\ER} = \sum_i \xi_i 
\left(\frac{d\mathcalR}{d\ER}\right)_i\,,
\label{eq:recoil-spect-i}
\end{equation}   
where $\xi_i$ is the mass fraction of the target element $i$ in the 
detector ($\sum_i \xi_i =1$), and the recoil rate 
$\left(d\mathcalR/d\ER\right)_i$ for the element 
$i$ is calculated from equation (\ref{eq:recoil-spect2}) with  
$A$ replaced by $A_i$ and $\mA$ by $\mAi$ in all the relevant 
quantities. 

The expected rate of events, $\mathcalR_{\rm exp}$, i.e., number of 
events per unit mass of the compound target material per unit time, is 
then given by 

\begin{equation}
\mathcalR_{\rm exp}= \sum_i \mathcalR_{\rm exp}^{(i)} = \sum_i \xi_i 
\int_{\ERthi}^{\ERmaxi} d\ER \,\epsilon_{i}(\ER) 
\left(\frac{d\mathcalR}{d\ER}\right)_i\,,
\label{eq:event-rate}
\end{equation}
where $\epsilon_{i}(\ER)$ is the bubble nucleation efficiency 
and $\ERthi$ is the recoil energy threshold for 
bubble nucleation by nuclei of element $i$, and  
\begin{equation}
\ERmaxi = \frac{2\mAi \vesc^2}{\left(1+\frac{\mAi}{\mchi}\right)^2}
\label{eq:ERmaxi}
\end{equation}
is the maximum recoil energy a nucleus of element $i$ can receive due to 
scattering with a WIMP of mass $\mchi$. Note that $\ERmaxi$ decreases 
with decreasing value of $\mchi$. Therefore, for a given target 
material element $i$, the condition $\ERmaxi  
\geq \ERthi$ for bubble nucleation by a recoiling nucleus of target 
element $i$ implies that the target element $i$ is insensitive 
to WIMPs of masses below a certain lowest value,  
$\mchilowest^{(i)}$, given by 
\begin{equation}
\mchilowest^{(i)} = \mAi 
\left[\left(\frac{2\mAi\vesc^2}{\ERthi}\right)^{1/2}-1\right]^{-1}\,. 
\label{eq:mchilowest}
\end{equation}
Note that, for a SLD, since $\ERthi$ depends on the operating 
temperature and pressure of the SLD as discussed in section 
\ref{sec:detector-principle} above, the lowest WIMP 
mass $\mchilowest^{(i)}$ that can be probed with target element $i$ 
also depends on the operating temperature and pressure of the SLD. 


\section{Results and Discussion}\label{sec:results}
\subsection{Threshold energies of recoiling \hydrogen1, \carbon12 and 
\fluorine19 nuclei for bubble nucleation in superheated liquid 
\c2h2f4}\label{subsec:results_Eths}
As discussed in section  \ref{sec:detector-principle}, to obtain the 
bubble nucleation threshold energy of a particular recoiling nucleus, we 
need to first compare the range of the nucleus at energy $\Ec/\etaT$ for 
a given value of $\etaT$ with the critical diameter $2\Rc$ at the given 
operating temperature and pressure of the SLD. We calculate the critical 
radius $\Rc$ [equation (\ref{eq:Rc_def})] and the critical 
energy $\Ec$ [equation (\ref{eq:Ec_def})] using values of the 
thermodynamic quantities taken from the REFPROP database maintained by 
the National Institute of Standards and Technology~\cite{REFPROP}. The 
ranges of \hydrogen1, \carbon12 and \fluorine19 nuclei in superheated 
liquid \c2h2f4 are calculated using the ``Stopping Range of Ions in 
Matter" (SRIM) software package~\cite{SRIM}. For simplicity, all 
results shown below are for operating pressure fixed at 1 atm. 

The values of $\Ec$ and $2\Rc$ are listed in Table 
\ref{table:Ec-Rc-Leff-Range} for various operating temperatures ranging from 
35$^\circ\,$C to 60$^\circ\,$C. The ranges ($R$) of  
\hydrogen1, \carbon12 and \fluorine19 nuclei of energy $\Ec$ in 
liquid \c2h2f4 at different operating temperatures are also listed in 
Table \ref{table:Ec-Rc-Leff-Range} for easy comparison with the 
values of critical diameter $2\Rc$ at the corresponding temperatures.     

\begin{table}
\smallskip
\centering
\begin{tabular}{|c|c|c|c|c|c|c|c|}
\hline
&     &  & &
\multicolumn{3}{c}{Range ($E=\Ec$)}\vline\\
\cline{5-7}
Operating & Critical & Critical & $2\Rc$
& \hydrogen1 & \carbon12 & \fluorine19\\
Temperature & Energy & Radius &&&&\\
($T$) & ($\Ec$) & ($\Rc$)  & (nm) & (nm) & (nm) & (nm)\\
($^{\circ}\,$C) & (keV) & (nm) & & & & \\

\hline
35 & 1.92 & 17.16  & 34.32 & 78.72 & 12.99 & 10.37\\
40 & 1.08 & 13.36  & 26.72 & 44.88 & 7.87  &  6.92\\
45 & 0.61 & 10.39  & 20.78 & 25.39 & 4.93  & 4.94\\
50 & 0.34 & 8.05   & 16.10 & 14.38 & 3.27  & 3.82\\
55 & 0.19 & 6.20   & 12.40 & 8.14  & 2.32  & 3.18\\
60 & 0.10 & 4.73   & 9.46  & 4.59  & 1.79  & 2.82\\
\hline
\end{tabular}
\caption{The critical energy $\Ec$, critical radius $\Rc$, 
and critical diameter $2\Rc$ for bubble nucleation in
superheated liquid \c2h2f4 at a pressure of 1 atm and
various operating temperatures. The values of the range, $R(E=\Ec)$, of
\hydrogen1, \carbon12 and \fluorine19 nuclei of energy $\Ec$ in liquid
\c2h2f4 at different temperatures are also listed for easy comparison
with the values of $2\Rc$ at the corresponding temperatures.}
\label{table:Ec-Rc-Leff-Range}
\end{table}

\begin{figure}[thp]
\centering
\includegraphics[width=\columnwidth]{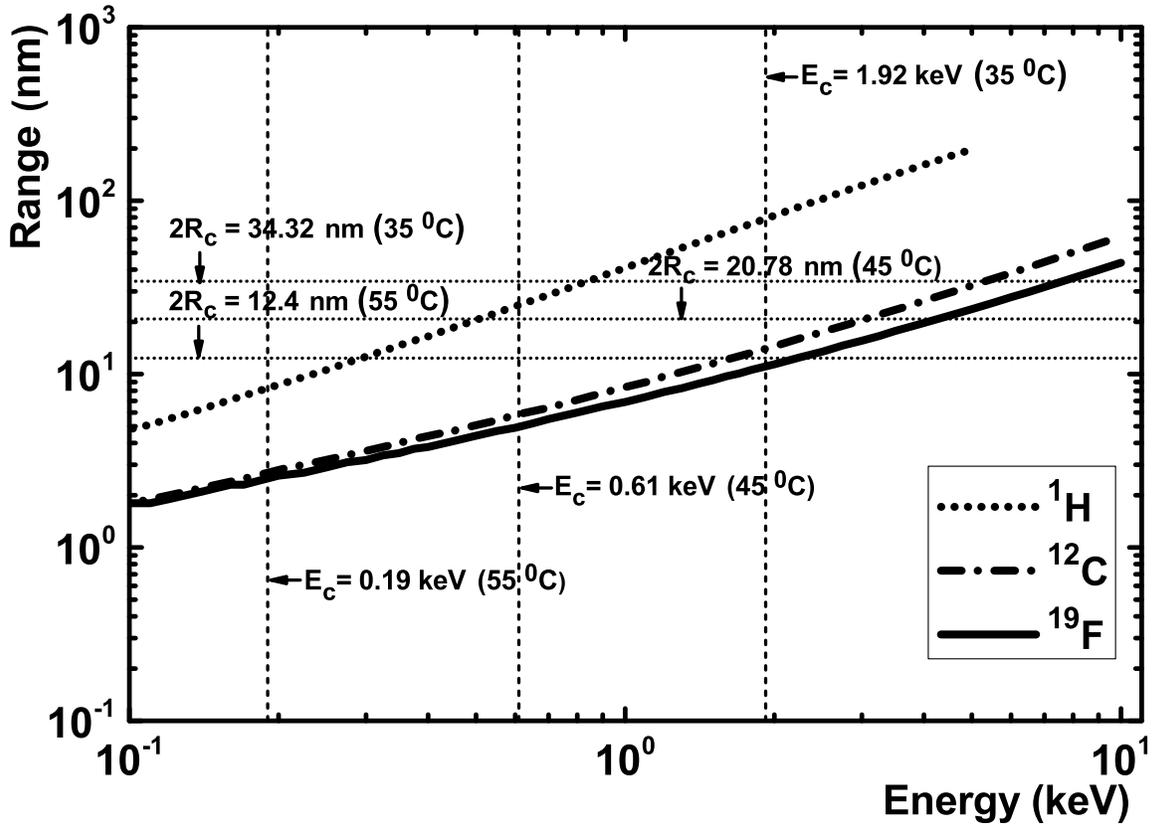}\\
\caption{Ranges of \hydrogen1, \carbon12 and \fluorine19 nuclei in 
superheated liquid \c2h2f4 as functions of their energy. The dashed   
vertical lines mark the critical energy $\Ec$ at three different 
temperatures, namely, $35^\circ\,$, $45^\circ\,$ and $55^\circ\,$C, and 
the dotted horizontal lines mark the corresponding values of 
$2\Rc$.} 
\label{fig:range-vs-energy_all-nuclei} 
\end{figure}

To aid visualization, we display in Figure 
\ref{fig:range-vs-energy_all-nuclei} the ranges of \hydrogen1, \carbon12 
and \fluorine19 nuclei as functions of their energy and their comparison 
with the critical diameter $2\Rc$ at three different temperatures, 
namely, $35^\circ\,$, $45^\circ\,$ and $55^\circ\,$C. 

It is seen that the ranges of \carbon12 and \fluorine19 at energy $\Ec$ 
at all temperatures of our interest are less than the corresponding 
critical diameter $2\Rc$. Thus, within the Seitz model, the threshold 
energies of these nuclei in the case of $\etaT=100\%$ are the same and 
equal to the Seitz threshold $\Ec$ at the corresponding temperature. 
For the case of $\etaT=50\%$, for example, the ranges of \carbon12 and 
\fluorine19 at the energy $2\Ec$ at all temperatures under consideration 
are also less than $2\Rc$, thus giving their bubble nucleation 
thresholds to be $2\Ec$, as expected, since in this case only 50\% of 
the total deposited energy goes into bubble nucleation. This behavior of 
nucleation thresholds of \carbon12 and \fluorine19 being set at 
$\Ec(T)/\etaT$ (owing to validity of the condition $R(\Ec(T)/\etaT)\leq 
2\Rc(T)$) remains true for lower values of 
$\etaT$ too, but only at progressively higher temperatures as the value 
of $\etaT$ is decreased.  

\begin{figure}[thp]
\centering
\includegraphics[width=0.8\columnwidth]{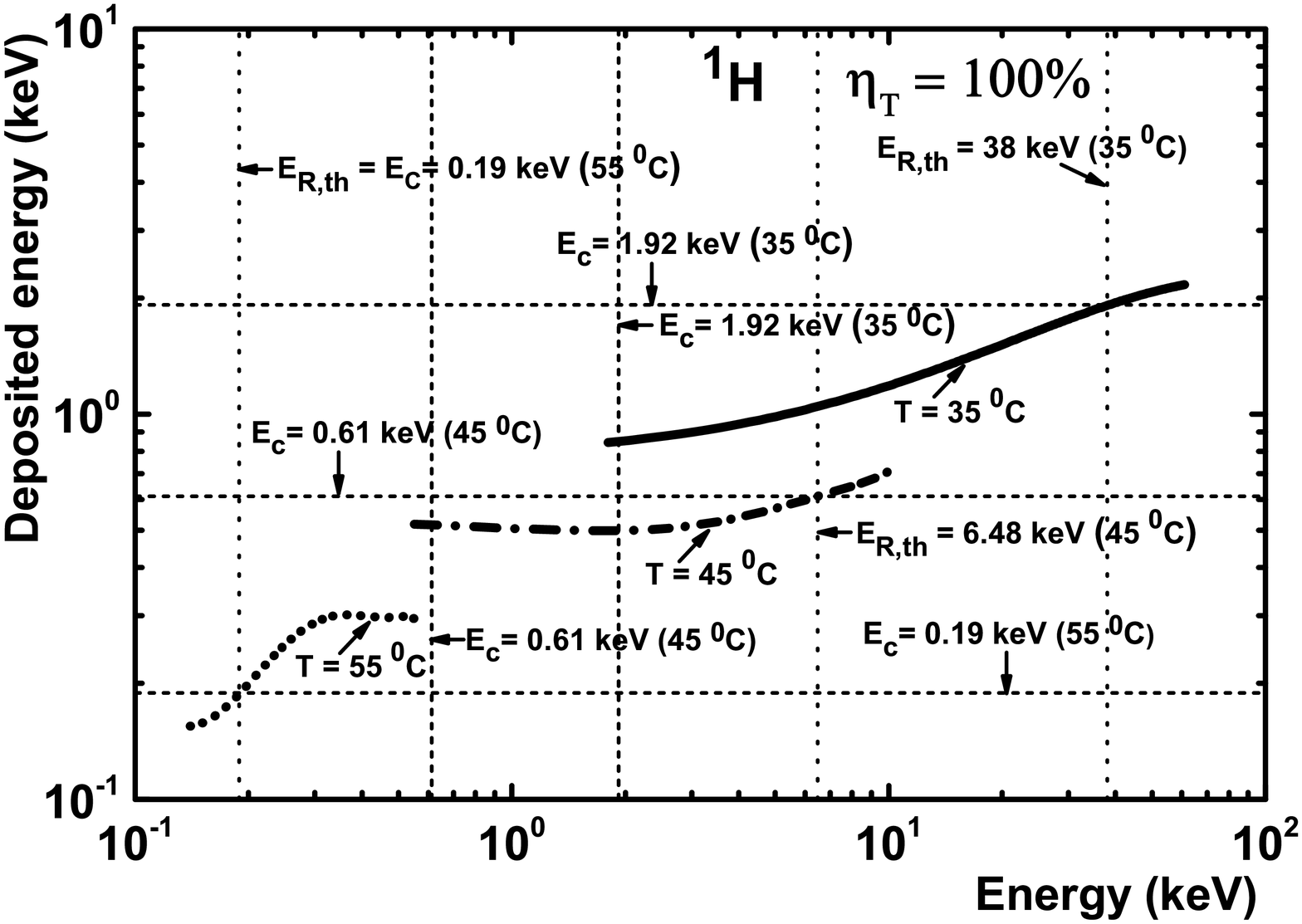}\\
\includegraphics[width=0.8\columnwidth]{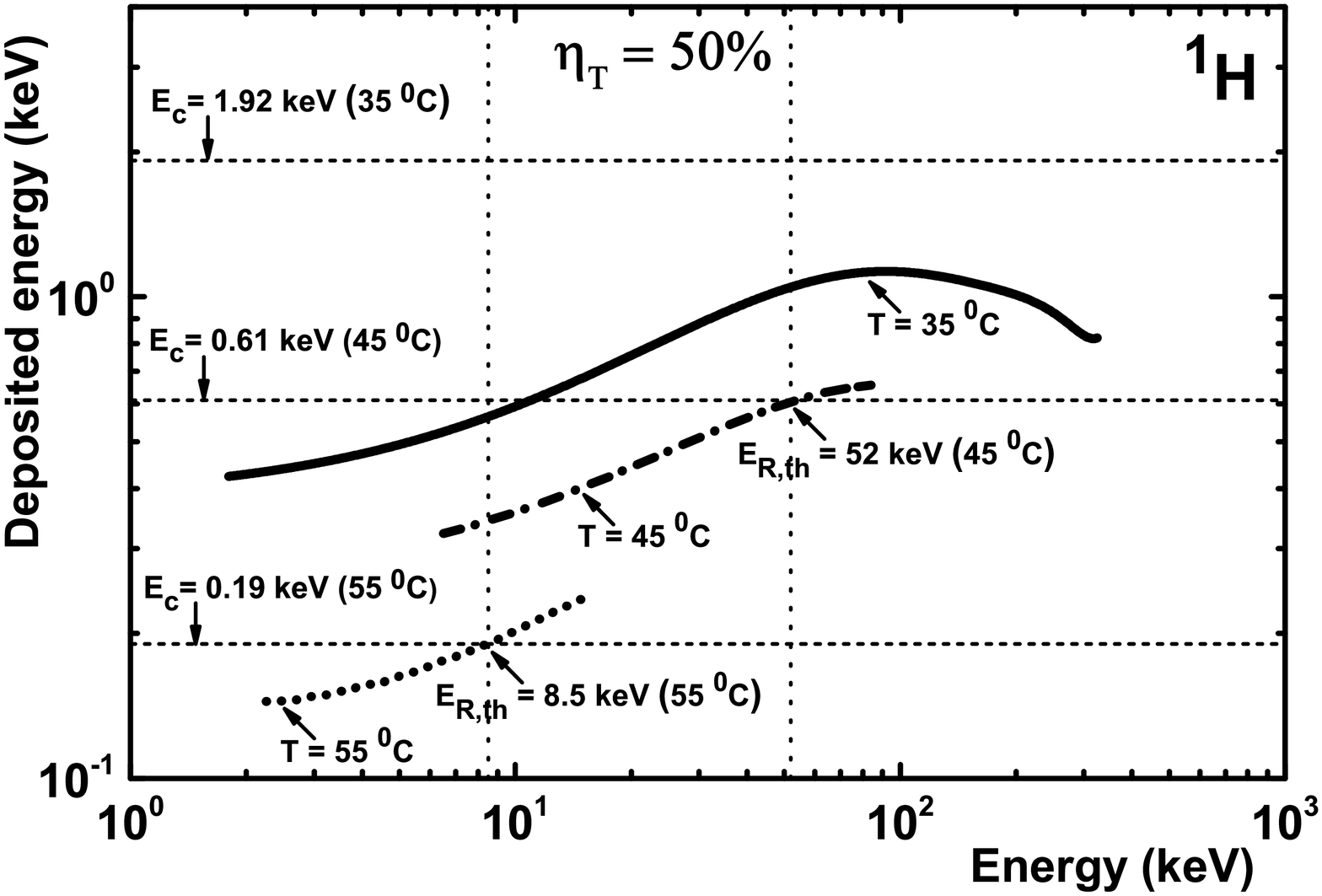}
\caption{Energy deposited by \hydrogen1 nuclei over the length scale 
$2\Rc$ in superheated liquid \c2h2f4 for temperatures 
$35^\circ\,$, $45^\circ\,$ and $55^\circ\,$C for $\etaT=100\%$ (upper 
panel) and $\etaT=50\%$ (lower panel). The vertical and 
horizontal dashed lines mark the values of the critical energy $\Ec$ at 
different temperatures. 
The \hydrogen1 threshold energies ($\ERth$) obtained from crossings of 
the energy deposition curves with the horizontal $\Ec$ lines for the 
three different temperatures are marked by vertical dotted lines. 
Note that for $55^\circ\,$C, $\ERth$ and $\Ec$ coincide in the case of 
$\etaT=100\%$ . Also, for $T=35^\circ\,$C, there is no solution of 
equation (\ref{eq:E_dep_condition}) for the case of $\etaT=50\%$. }      
\label{fig:Edep-vs-E_H_b2} 
\end{figure}

\begin{figure}[thp]
\centering
\includegraphics[width=0.7\columnwidth]{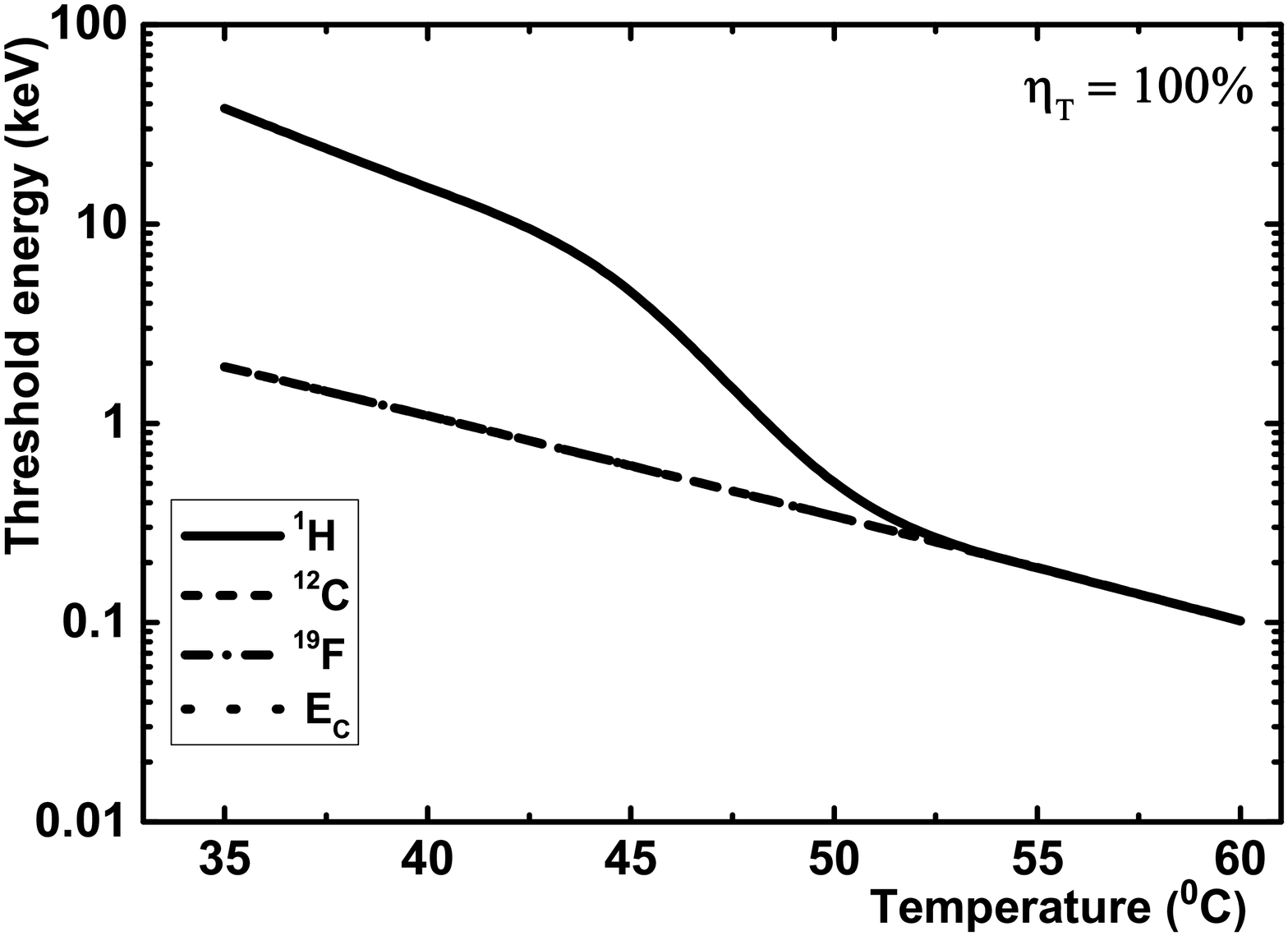}\\
\includegraphics[width=0.7\columnwidth]{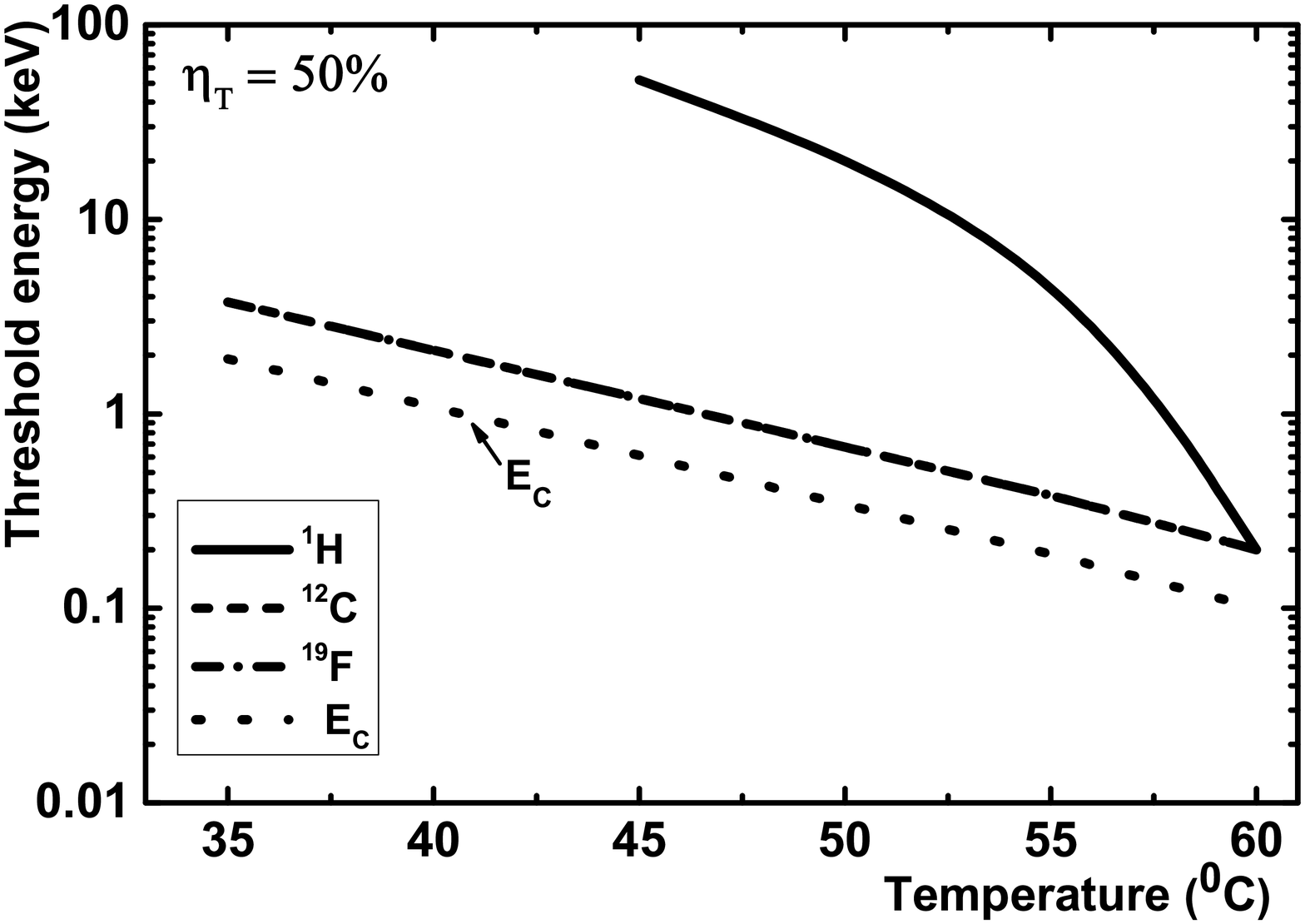}
\caption{The threshold energy for bubble nucleation by 
recoiling \hydrogen1, \carbon12 and \fluorine19 for the case $\etaT = 
100\%$ (upper panel) and $\etaT = 50\%$ (lower panel) in superheated 
liquid \c2h2f4 as functions of temperature. Note that the thresholds of 
\carbon12 and \fluorine19 coincide with $\Ec(T)$ for $\etaT=100\%$, and 
with $2\Ec(T)$ for the case $\etaT=50\%$. }
\label{fig:ERth-vs-temp}
\end{figure}

The case of \hydrogen1 is, however, very different. The range of 
\hydrogen1 at $\Ec$ is larger than $2\Rc$ at all 
temperatures below $\sim 50^\circ\,$C. Thus, even in the case of 
$\etaT=100\%$, the bubble nucleation threshold energy of \hydrogen1 at 
temperatures below $\sim 50^\circ\,$C will be larger than $\Ec$ at the 
corresponding temperatures and will have to be determined by finding the 
energy at which equation 
(\ref{eq:E_dep_condition}) is satisfied. This is illustrated in Figure 
\ref{fig:Edep-vs-E_H_b2} for $\etaT=100\%$ and 50\%.   
The resulting threshold energies of \hydrogen1, \carbon12 and 
\fluorine19 as functions of temperature are shown graphically in Figure 
\ref{fig:ERth-vs-temp} for the two cases of $\etaT=100\%$ and 50\%. 

\begin{figure}[thp]
\centering
\includegraphics[width=0.7\columnwidth]{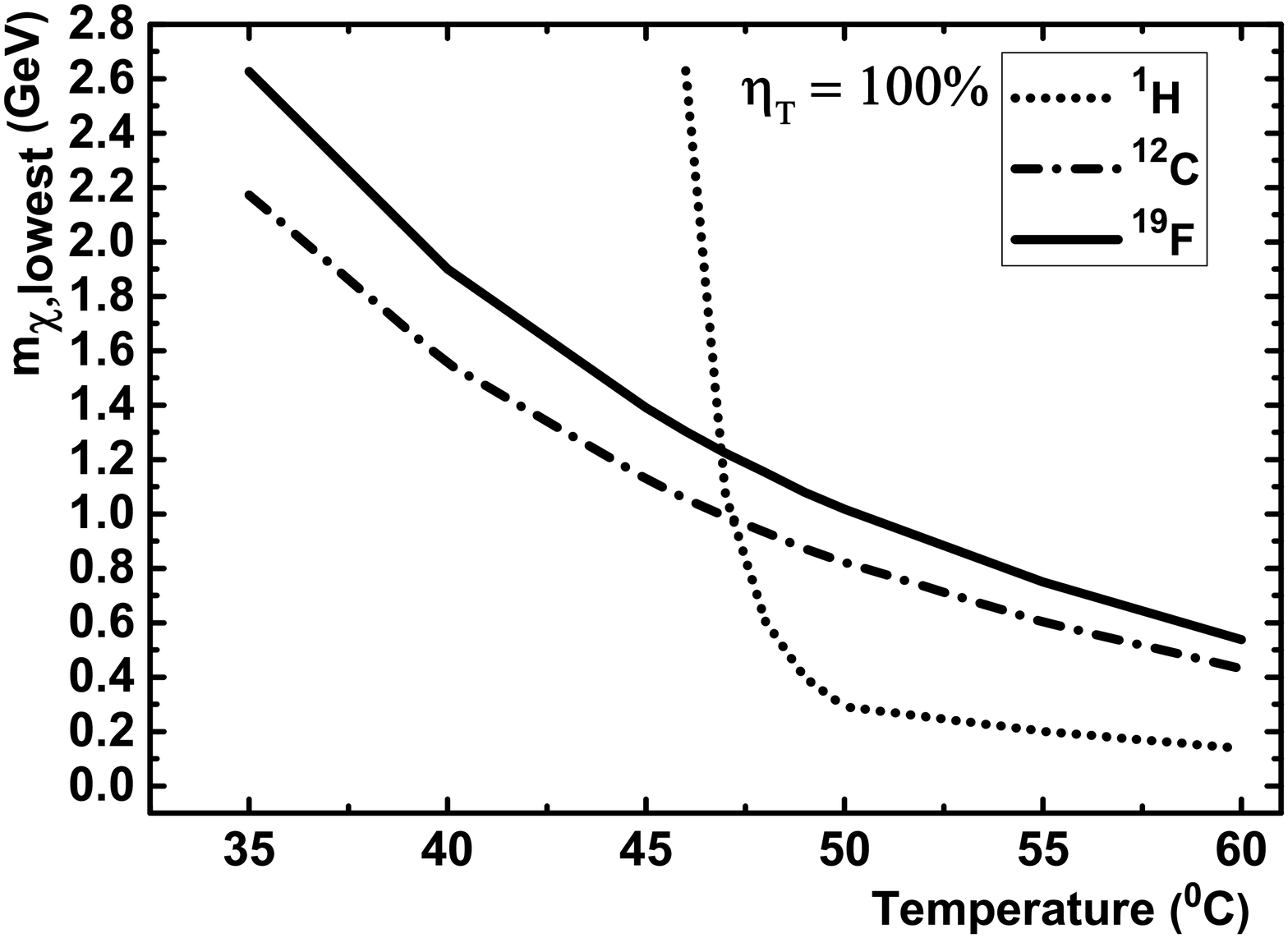}\\
\includegraphics[width=0.7\columnwidth]{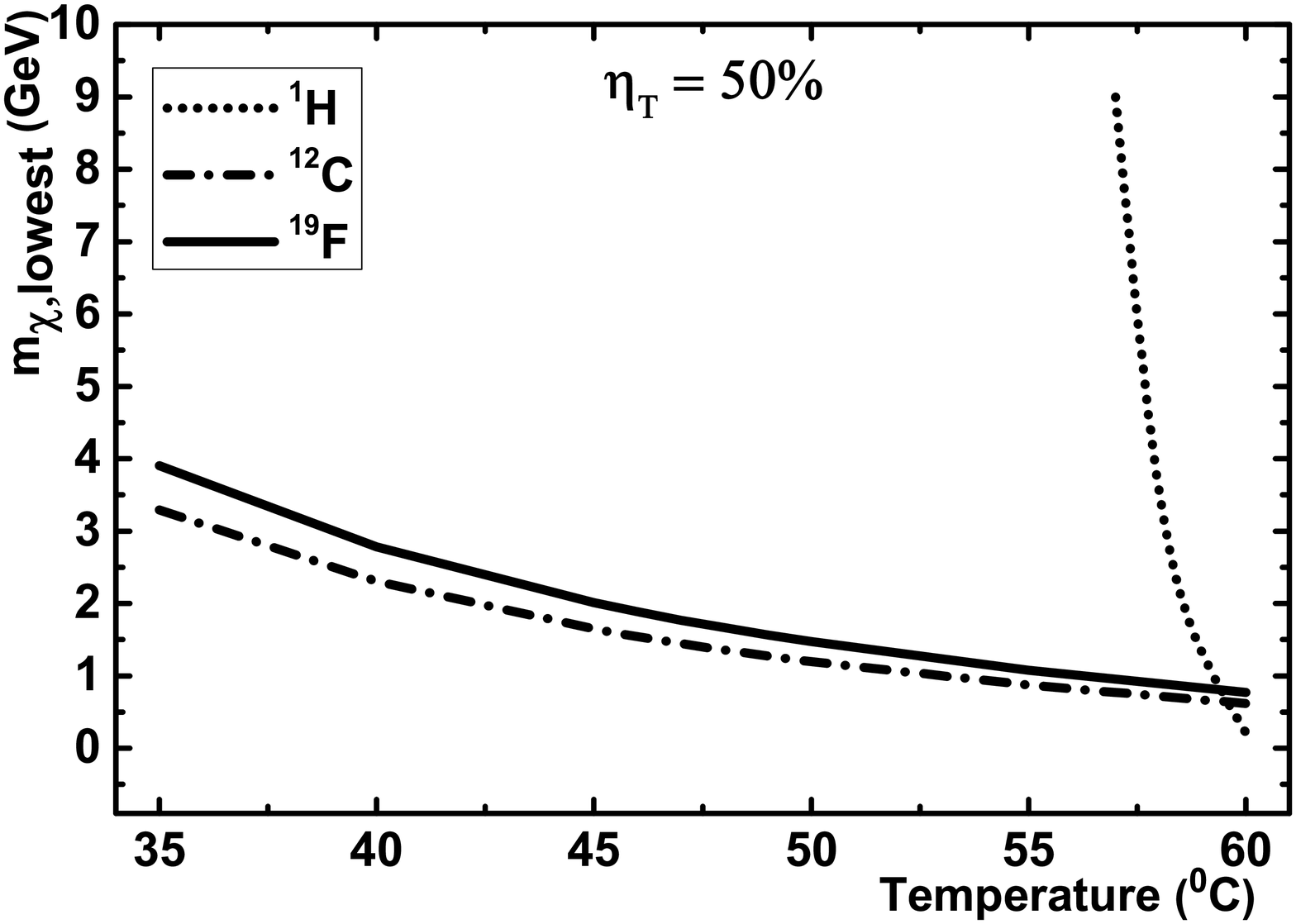}
\caption{Lowest WIMP masses that can produce recoiling \hydrogen1,
\carbon12 and \fluorine19 nuclei above their bubble
nucleation threshold energies (see Table
\ref{table:Eth-mchilowest-vs-temp}) in superheated liquid \c2h2f4 as a
function of temperature, for $\eta_T = 100\%$ (top panel) and $\eta_T =
50\%$ (bottom panel).}
\label{fig:mchilowest-vs-temp_vesc-540}
\end{figure}

With the threshold energies of \hydrogen1, \carbon12 and \fluorine19 
determined as above, we can calculate the temperature dependence 
of the lowest WIMP mass to which superheated liquid \c2h2f4 can be 
sensitive for different values of $\etaT$, using equation 
(\ref{eq:mchilowest}). The results are
displayed in Table \ref{table:Eth-mchilowest-vs-temp} and shown
graphically in Figure \ref{fig:mchilowest-vs-temp_vesc-540}.

\begin{table}
        \smallskip
        \centering
\begin{tabular}{|c|c|c|c|c|c|c|c|c|c|c|c|c|c|c|c|c|c|c|}
        \hline

   &   \multicolumn{6}{c}{($\eta_T$ = 25\%)}\vline &
        \multicolumn{6}{c}{($\eta_T$ = 50\%)}\vline & 
\multicolumn{6}{c}{($\eta_T$ = 100\%)}\vline\\

        \cline{2-7}\cline{7-13}\cline{13-19}
  T     & \multicolumn{3}{c}{$\ERth$}\vline &
        \multicolumn{3}{c}{$\mchilowest$}\vline &
        \multicolumn{3}{c}{$\ERth$}\vline &
        \multicolumn{3}{c}{$\mchilowest$}\vline &
        \multicolumn{3}{c}{$\ERth$}\vline &
        \multicolumn{3}{c}{$\mchilowest$}\vline\\
        & \multicolumn{3}{c}{(keV)}\vline &
        \multicolumn{3}{c}{(GeV)}\vline &
        \multicolumn{3}{c}{(keV)}\vline &
        \multicolumn{3}{c}{(GeV)}\vline &
        \multicolumn{3}{c}{(keV)}\vline &
        \multicolumn{3}{c}{(GeV)}\vline\\
        
\cline{2-4}\cline{5-7}\cline{8-10}\cline{11-13}\cline{14-16}\cline{17-19}
        $(^{\circ}\,$C) & \hydrogen1 & \carbon12 & \fluorine19 & 
\hydrogen1 &
        \carbon12 &  \fluorine19 & \hydrogen1 & \carbon12 & \fluorine19 
& \hydrogen1 &
        \carbon12 & \fluorine19 & \hydrogen1 & \carbon12 & \fluorine19 & 
\hydrogen1 & \carbon12 & \fluorine19\\

        \hline
        35 & - & 250.0 & 8.0 & - & - & 6.34 & -  & 3.84  & 3.84 & - & 
3.29 & 3.90 & 38.0 & 1.92 & 1.92 & - & 2.17 & 2.63\\
        40 & - &  22.0 & 4.32 & - & 13.69 & 4.25 & - & 2.16 & 2.16 & - & 
2.36 & 2.84 & 15.16 & 1.08 & 1.08 & - & 1.55 & 1.90\\
        45 & - & 2.44 & 2.44 & - & 2.54 & 3.06 &  52.0 & 1.22 & 1.22  & 
-  & 1.65 & 2.01 & 6.48 & 0.61 & 0.61 & - & 1.13 & 1.39\\
        50 & - & 1.36 & 1.36 & - & 1.77 & 2.16 & 20.0 & 0.68 & 0.68 & - 
& 1.20 & 1.48 & 0.34 & 0.34 & 0.34 & 0.29 & 0.82 & 1.02\\
        55 & 55.0 & 0.76 & 0.76 & -  & 1.31 & 1.61 & 8.50 & 0.38 & 0.38 
& - & 0.87 & 1.08 & 0.19 & 0.19 & 0.19 & 0.20 & 0.60 & 0.75\\
        60 & 22.50 & 0.40 & 0.40 & - & 0.89 & 1.11 & 0.20 & 0.20 & 0.20 
& 0.21 & 0.62 & 0.77 & 0.10 & 0.10 & 0.10 & 0.14 & 0.43 & 0.54\\
   \hline
\end{tabular}
\caption{Threshold energies of WIMP-induced recoiling \hydrogen1,
\carbon12 and \fluorine19 nuclei for bubble nucleation in superheated 
liquid \c2h2f4 at different temperatures for $\eta_T = 25\%$ , $\eta_T = 
50\%$ and $\eta_T = 100\%$ and the corresponding lowest values of the 
WIMP mass that can produce those recoil nuclei of the respective 
threshold energies, i.e., the lowest mass WIMPs that can 
be probed with a \c2h2f4 superheated liquid detector. A blank (-) entry 
indicates no sensitivity to the target element at the temperature 
under consideration, i.e., $\ERthi (T) > \ERmaxi$ for the target element 
$i$.}
\label{table:Eth-mchilowest-vs-temp}
\end{table}

From Table \ref{table:Eth-mchilowest-vs-temp} and Figure
\ref{fig:mchilowest-vs-temp_vesc-540}, we see that, with suitable 
choice of the operating temperature, a \c2h2f4 SLD can 
serve as a good detector for very low mass (sub-GeV -- few GeV) WIMPs. 
In a separate experimental work~\cite{c2h2f4-SunitaSahoo-2019} 
it is shown that at temperatures $T < (38.5 \pm 1.4)^\circ\,$C, 
\c2h2f4 is 
insensitive to gamma rays (which can cause nucleation events 
through electron recoils), though sensitive to neutrons (which give  
nucleation events through nuclear recoils). Thus, a sufficiently 
large \c2h2f4 SLD operated 
at $T\sim 35^\circ\,$C, for example, would be sensitive to 
WIMP-induced \carbon12 recoil events for WIMPs of mass in the 
few GeV range down to $\sim$ 2.2 GeV in the case of $\etaT=100\%$ and 
$\sim$ 3.3 GeV in the case of $\etaT=50\%$ without 
being sensitive to background $\gamma$-rays. However, for sub-GeV mass 
WIMPs, the SLD would need to be operated at higher temperatures. 
From Table \ref{table:Eth-mchilowest-vs-temp} and Figure
\ref{fig:mchilowest-vs-temp_vesc-540}, we see that, the presence of 
hydrogen in \c2h2f4 can make the SLD sensitive to WIMPs of mass down to 
$\lsim$ 200 MeV at temperatures $T\gsim 60^\circ\,$C for $\etaT\geq 
50\%$. However, at these 
temperatures the \c2h2f4 SLD becomes sensitive to background $\gamma$-rays 
as well.  

Note from equation (\ref{eq:mchilowest}) that  for a given 
recoiling nucleus $i$, $\mchilowest^{(i)}$ roughly scales as 
$({\ERthi})^{1/2}$ for $\ERthi \ll 2\mAi\vesc^2$. Thus, for \carbon12 
and \fluorine19, an upward shift of the bubble nucleation threshold 
energy from the Seitz threshold $\Ec$ by a factor of 2 (as is the case 
for \carbon12 and \fluorine19 when $\etaT$ changes from 100\% to 50\%), 
for example, shifts the corresponding $\mchilowest$ upwards by a factor 
roughly between 1.4 and 1.5. So, as far as probing the few-GeV WIMP-mass 
region with \carbon12 and \fluorine19 recoils in \c2h2f4 is concerned, 
we do not expect substantial changes in our results for the 
WIMP-mass sensitivity of \c2h2f4 estimated assuming the relevant 
thresholds fixed at the respective Seitz thresholds $\Ec$ (i.e., 
assuming $\etaT=100\%$) unless the actual threshold energies of these 
nuclei are larger than their Seitz thresholds by factors substantially 
larger than 2 (i.e., $\etaT\ll50\%$).         

\subsection{Recoil spectra and event rates for low mass WIMPs}
\label{subsec:recoil-spectra-and-event-rates} 
\begin{figure}[thp]
\centering
\includegraphics[width=0.5\columnwidth]{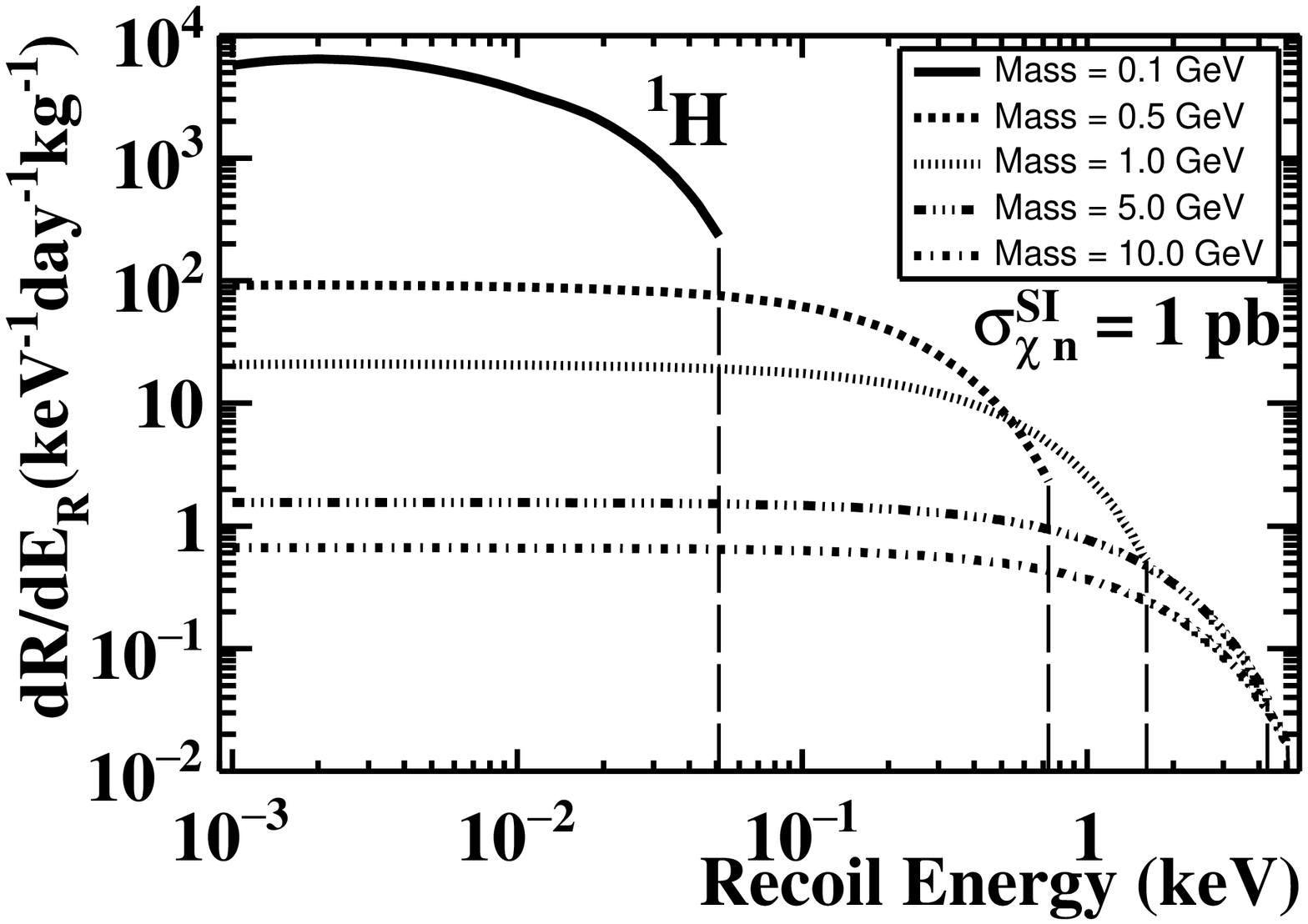}\\
\includegraphics[width=0.5\columnwidth]{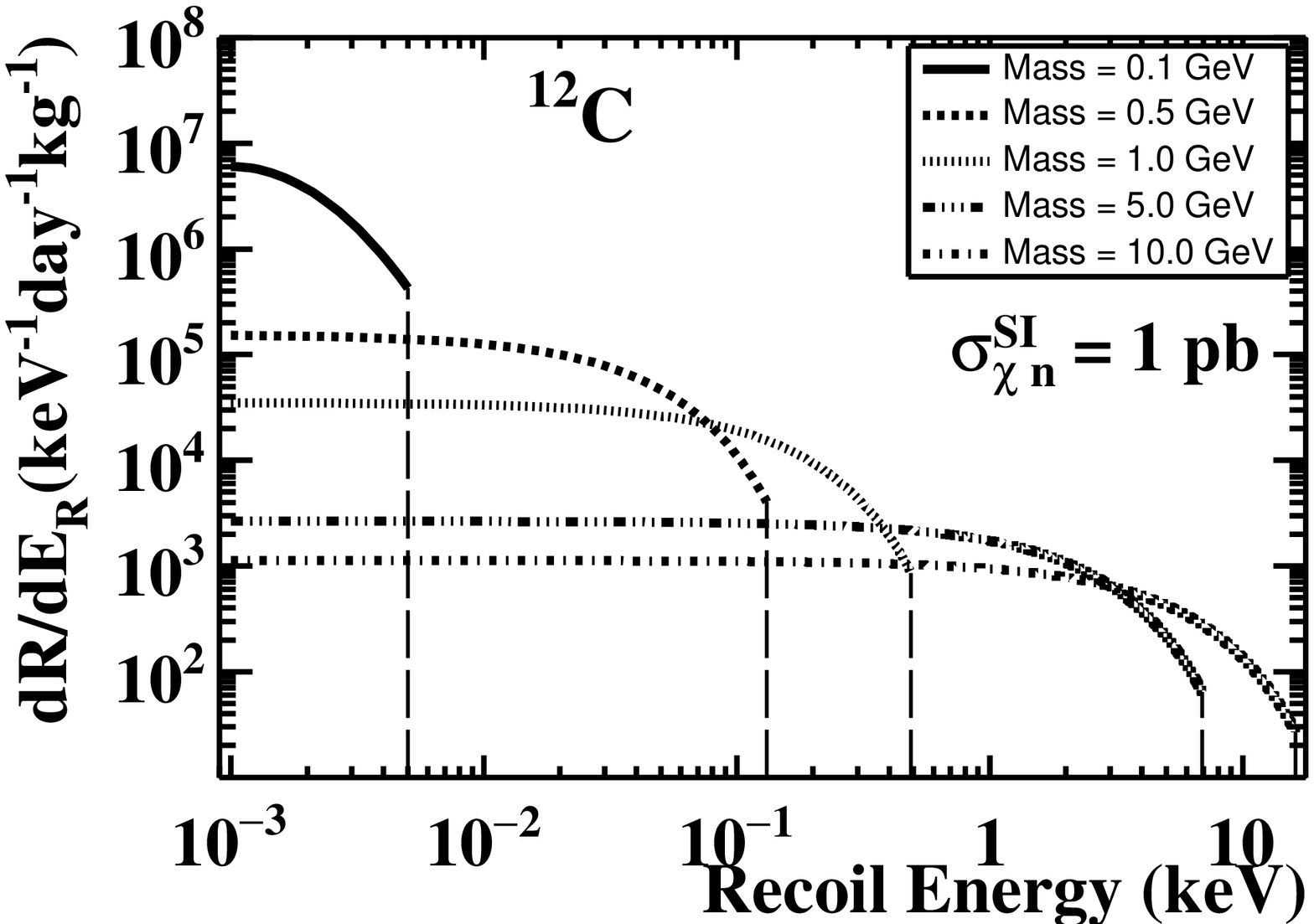}\\
\includegraphics[width=0.5\columnwidth]{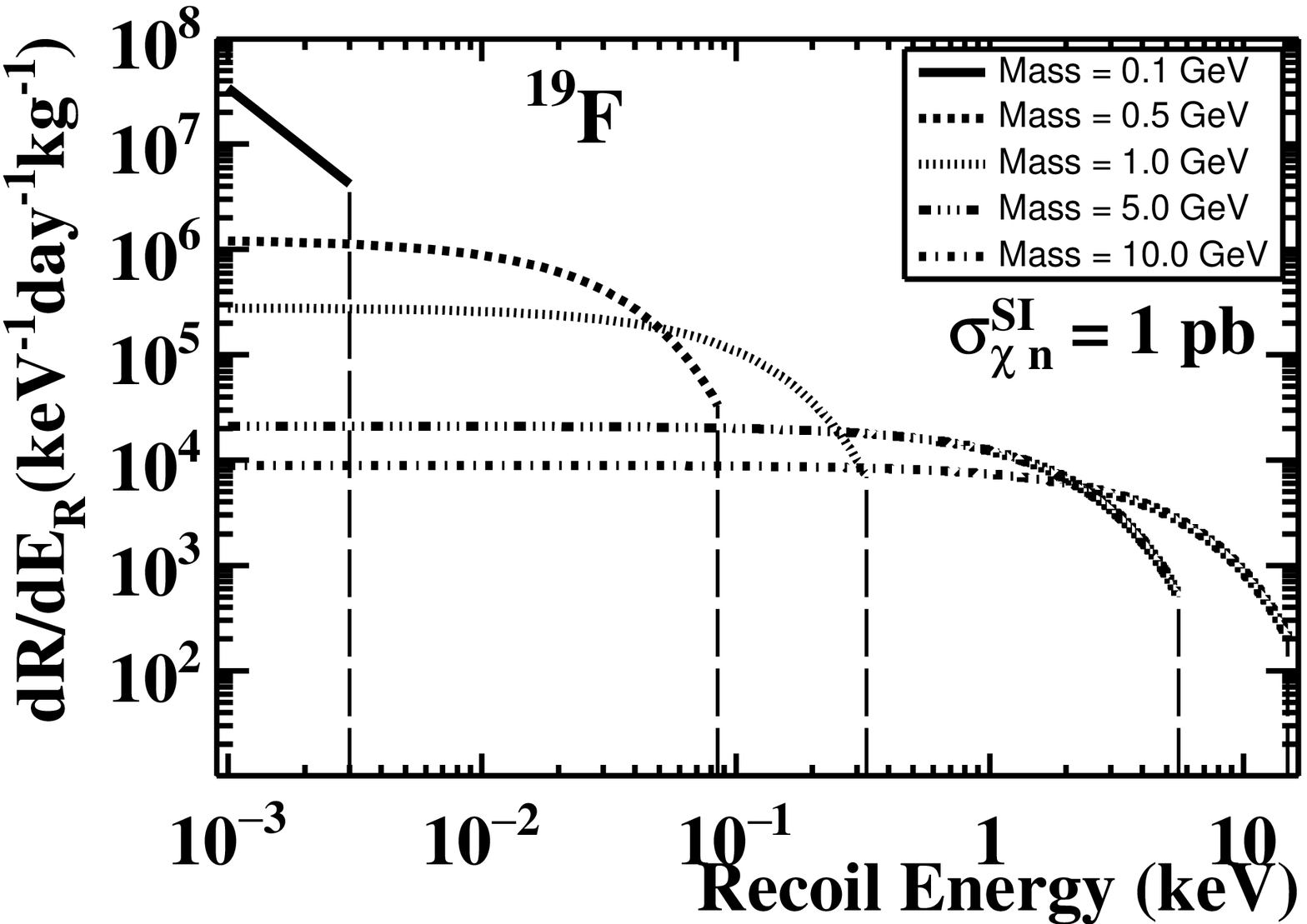}\\
\caption{Differential recoil energy spectra, as
given by equation (\ref{eq:recoil-spect2}), of
\hydrogen1, \carbon12 and \fluorine19 nuclei in \c2h2f4 (per keV of 
recoil energy per day per kg of \c2h2f4) for
different WIMP masses for a benchmark value of the WIMP-nucleon 
spin-independent (SI) cross section, 
$\sigmachinSI=1\pb\, (=10^{-36}\cm^2)$. (Note: One kg of \c2h2f4 
contains 0.02 kg of \hydrogen1, 0.235 kg of \carbon12 and
0.745 kg of \fluorine19.) Other parameter values used are: $\rhodm =
0.3\GeV/\cm^3$, $v_0=220\kmps$, $\vE=232\kmps$ and 
$\vesc=540\kmps$. The dashed vertical lines mark the sharp cutoff of the 
recoil spectra due to the sharp cutoff of the speed distribution of the 
WIMPs at the escape velocity $\vesc$ (see text).}
\label{fig:RecoilSpecs}
\end{figure}

The low WIMP-mass sensitivity of a \c2h2f4 SLD can be seen more clearly 
by looking at the expected contributions of \hydrogen1, 
\carbon12 and \fluorine19 nuclei to the rate of WIMP-induced nuclear 
recoil events. For this purpose, we calculate the differential recoil 
spectra of \hydrogen1, \carbon12 and \fluorine19 nuclei in \c2h2f4 from 
equation (\ref{eq:recoil-spect2}) for a benchmark value of 
$\sigmachinSI=1\pb$ --- these are shown in Figure 
\ref{fig:RecoilSpecs} --- and then integrate these spectra (see equation 
(\ref{eq:event-rate})) over the recoil energies to obtain the rates as a 
function of WIMP mass at various temperatures using the 
corresponding threshold energies listed in Table 
\ref{table:Eth-mchilowest-vs-temp}. 

\begin{figure}[thp]
\centering
\includegraphics[width=0.7\columnwidth]{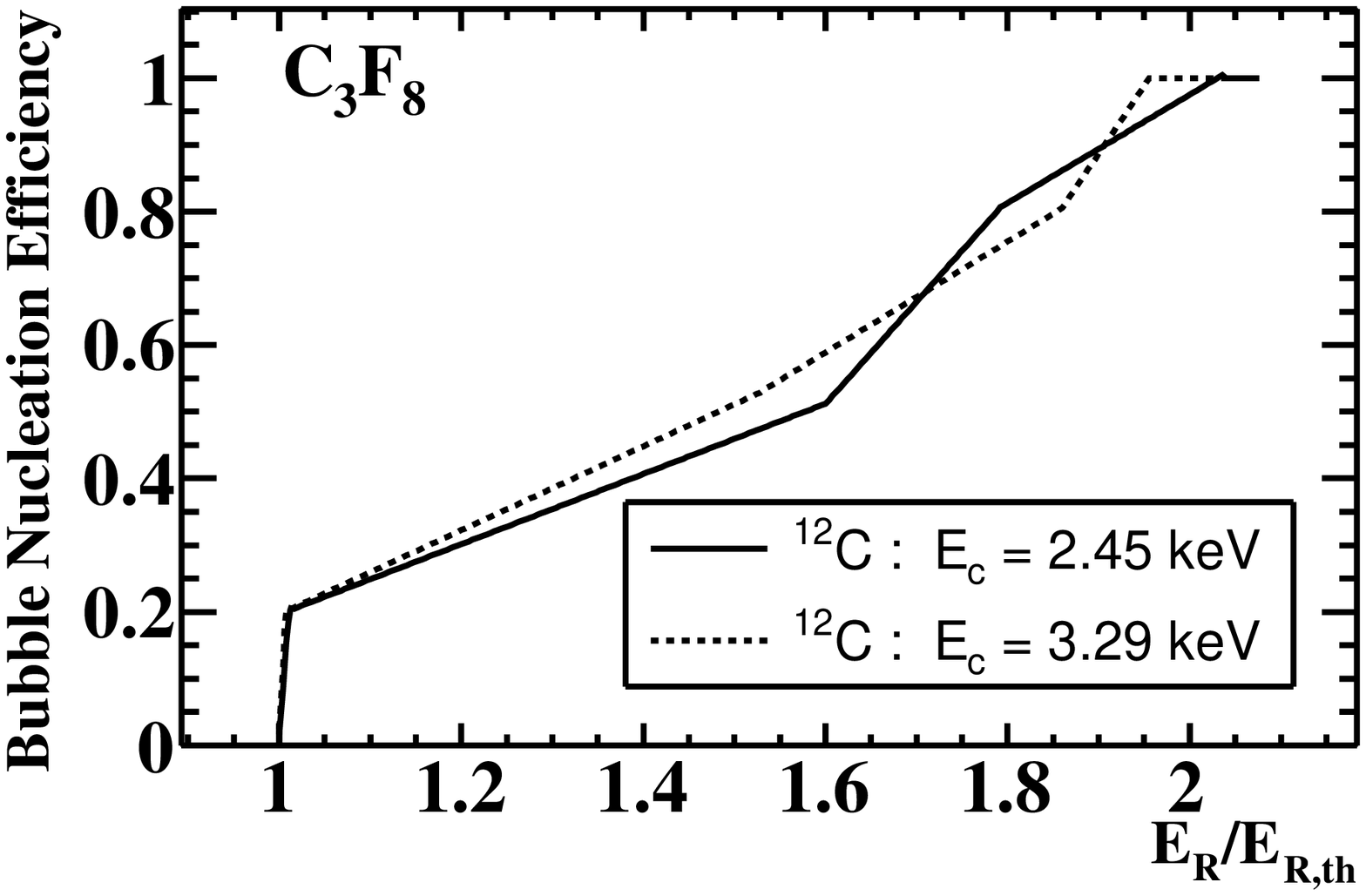}\\
\includegraphics[width=0.7\columnwidth]{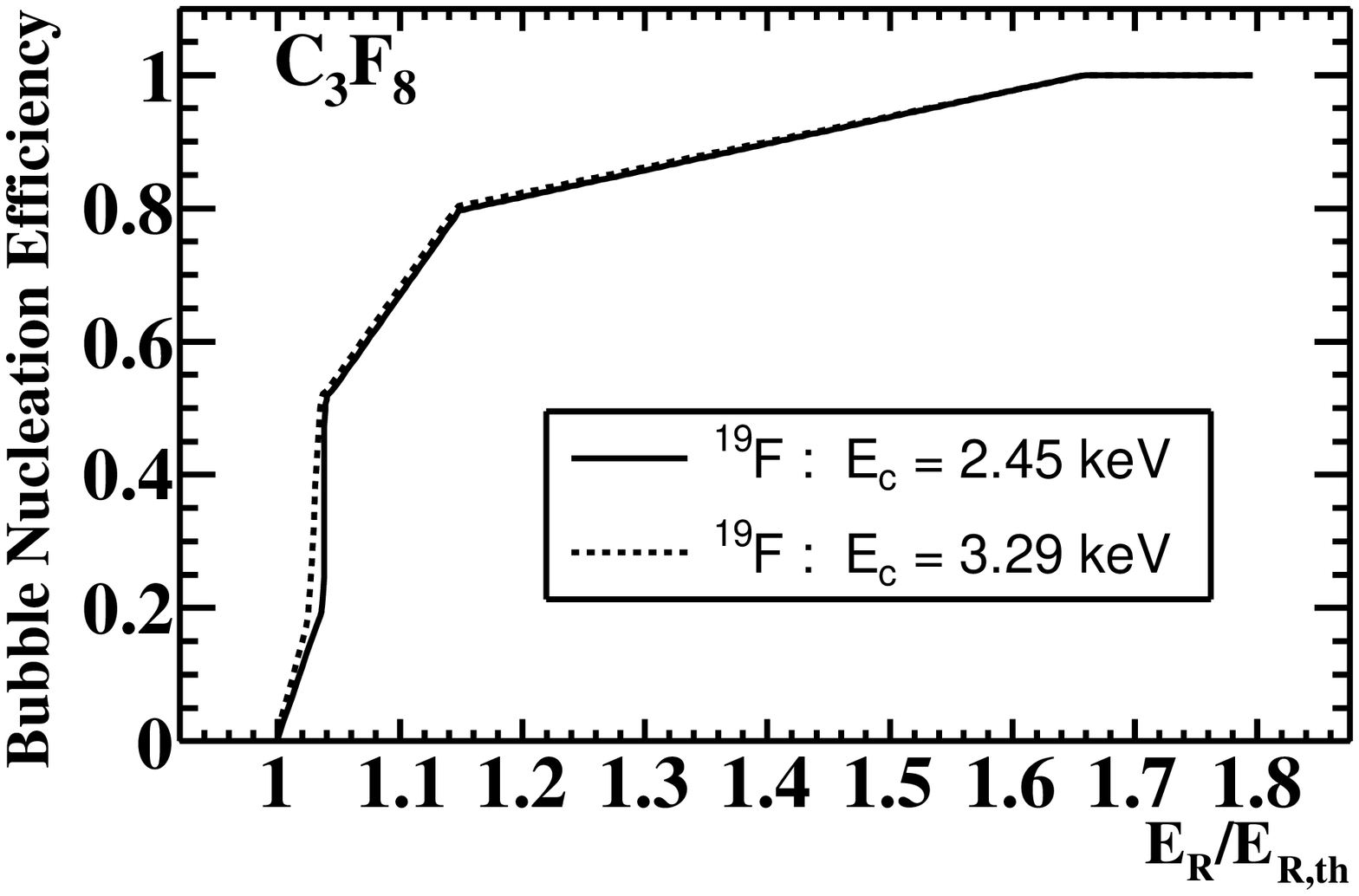}\\
\caption{Piecewise linear function fits to the best-fit bubble 
nucleation efficiency curves for \carbon12 (upper panel) 
and \fluorine19 (lower panel) recoils in \cthreef8 taken from Figure 3 
of Ref.~\cite{pico-60-C3F8-PRD2019}. In each case, the curves 
reconstructed from the upper as well as 
lower panels of Figure 3 of 
Ref.~\cite{pico-60-C3F8-PRD2019} corresponding to two different values 
of the Seitz thresholds $\Ec$=2.45 (solid curves) and 3.29 keV (dashed 
curves) in \cthreef8, 
respectively, are shown for comparison.}
\label{fig:eff-curves} 
\end{figure}

Following the discussions in section 
\ref{sec:detector-principle}, we take the 
bubble nucleation efficiencies, $\epsilon_i(\ER)$, in equation  
(\ref{eq:event-rate}) for the recoiling \carbon12 and \fluorine19 nuclei 
in \c2h2f4 to be the same as the respective efficiencies for these 
nuclei in the liquid \cthreef8 determined by the PICO 
experiment~\cite{pico-60-C3F8-PRD2019}. The piecewise linear function 
fits to the best-fit efficiency curves 
for \carbon12 and \fluorine19 
taken from the respective curves in the upper and lower panels of Figure 
3 of Ref.~\cite{pico-60-C3F8-PRD2019} are shown in Figure 
\ref{fig:eff-curves}. 
We see some small differences in the shapes of the 
efficiency curves for \carbon12  
obtained from the upper and lower panels of Figure 
3 of Ref.~\cite{pico-60-C3F8-PRD2019} (which correspond to the solid and 
dashed curves, respectively, in our Figure \ref{fig:eff-curves}), while 
the two curves for \fluorine19  are essentially identical. We verify 
that these differences between the solid and dashed curves for  
\carbon12 in Figure \ref{fig:eff-curves} do not make 
any significant difference in our results for the event rates and 
WIMP-mass sensitivities presented below. For definiteness, 
below we shall use the efficiency curves for \carbon12 
and \fluorine19 recoils represented by the solid curves in the 
upper and lower panels, respectively, of Figure \ref{fig:eff-curves}, 
for calculating the event rates and WIMP-mass sensitivities due to 
\carbon12 and \fluorine19 recoils in \c2h2f4. For \hydrogen1 
recoils, as already mentioned, we shall use a step-function efficiency 
curve at the relevant bubble nucleation threshold for \hydrogen1.     

\begin{figure}[thp]
\centering
\includegraphics[width=0.7\columnwidth]{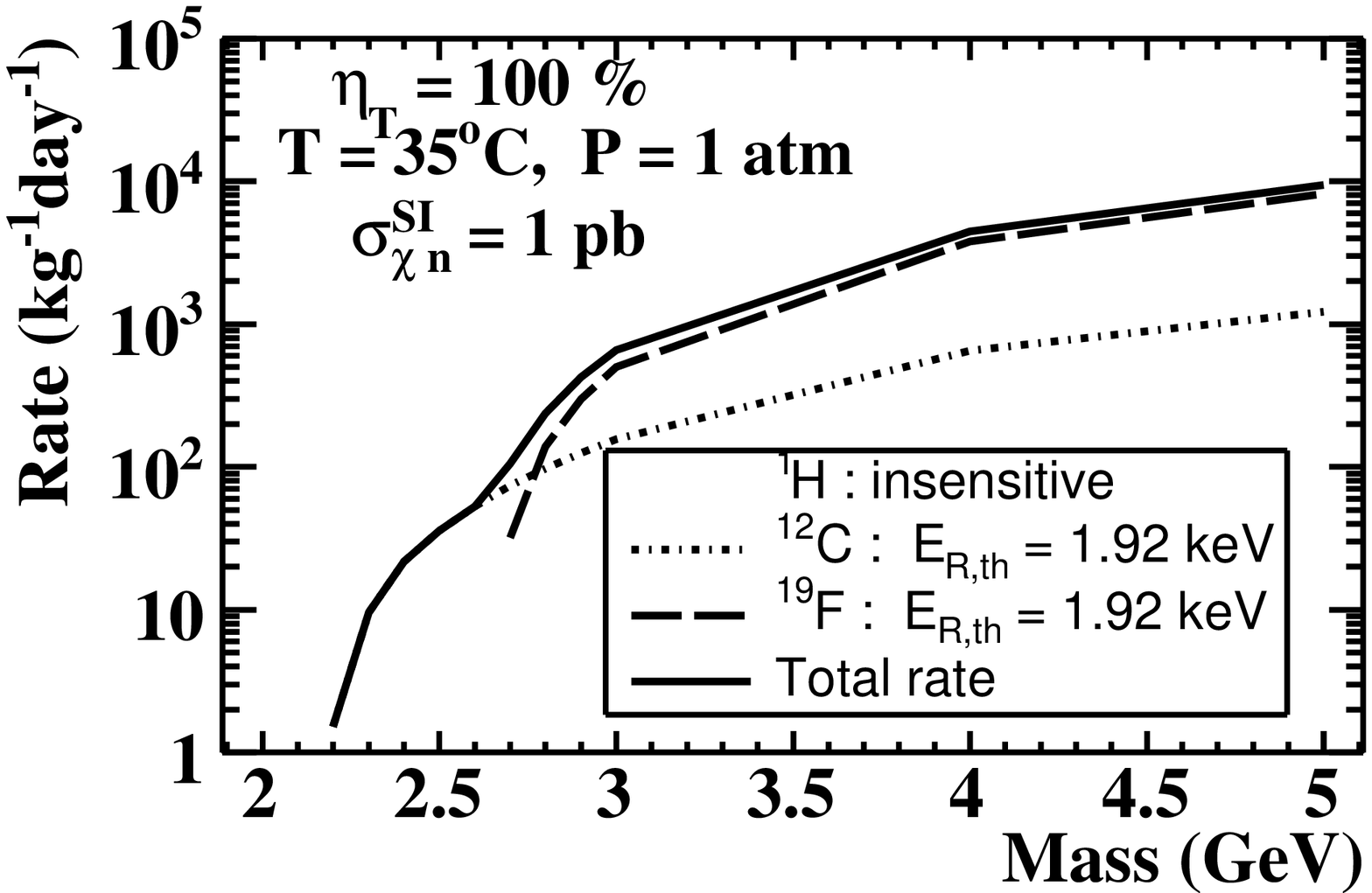}\\
\includegraphics[width=0.7\columnwidth]{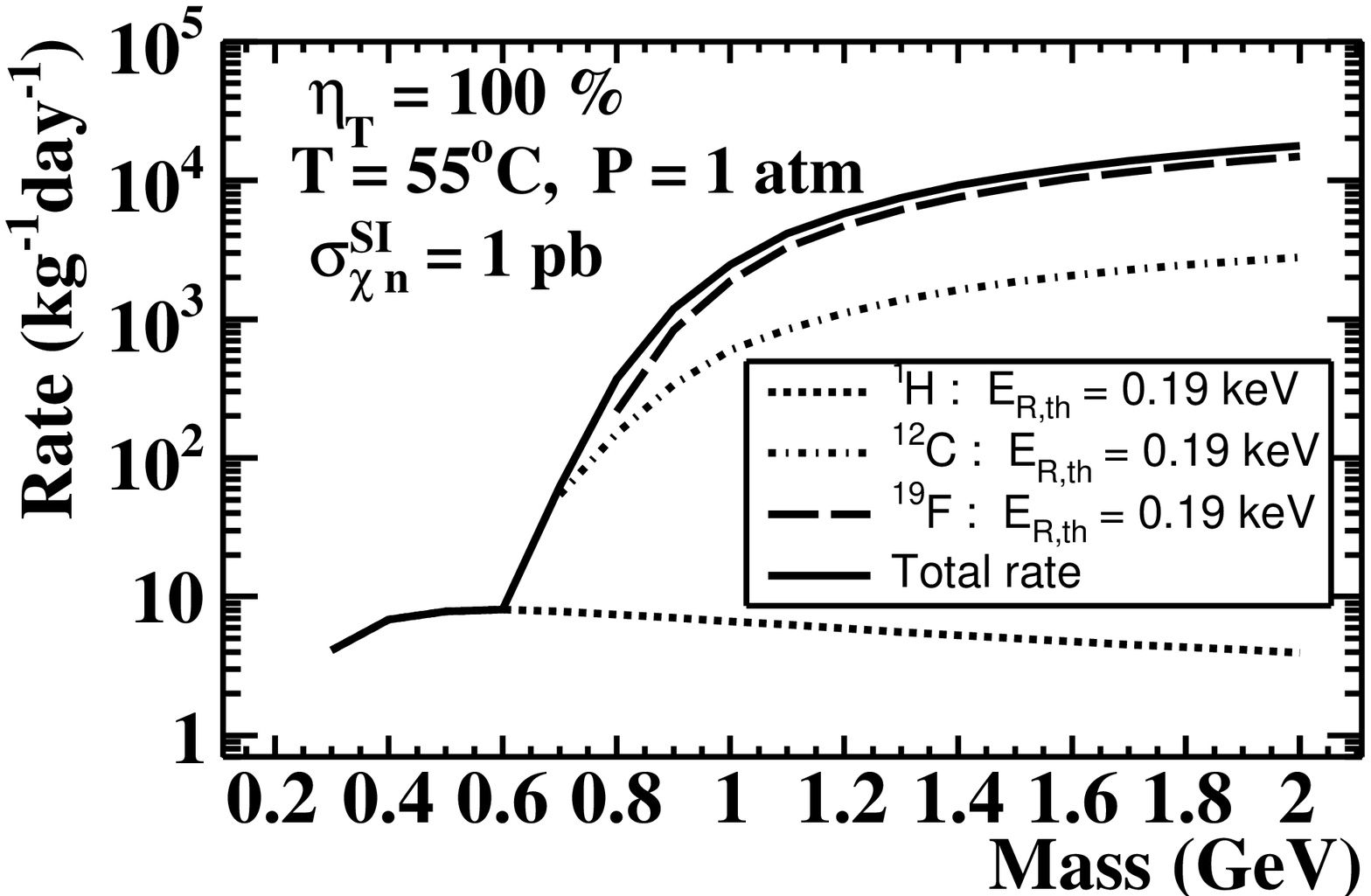}\\
\caption{Contributions of 
\hydrogen1, \carbon12 and \fluorine19 to the 
total rate of WIMP-induced nuclear recoil events in a \c2h2f4 SLD  
operated at a temperature of $35^\circ\,$C (upper panel) and 
$55^\circ\,$C (lower panel) and pressure of 1 atm, as a 
function of the WIMP mass ($\protect\lsim$ few GeV) for a benchmark 
value of spin-independent WIMP-nucleon cross section, 
$\sigmachinSI=1\pb$, for the case of $\etaT=100\%$. 
The bubble nucleation thresholds of individual nuclei used in the 
calculation are as listed in Table \ref{table:Eth-mchilowest-vs-temp}). 
Note that, at 
$T=35^\circ\,$C, the WIMP-induced recoiling \hydrogen1 
nuclei do not contribute any event since their maximum 
possible recoil energies (see Figure \ref{fig:RecoilSpecs}) are below 
their bubble-nucleation threshold energy at this temperature; see 
Figure \ref{fig:ERth-vs-temp} and Table 
\ref{table:Eth-mchilowest-vs-temp}. The sharp fall-off of the rates for  
different nuclear species at the lower mass end reflects the lowest 
WIMP mass sensitivities of the different nuclei shown in   
Table \ref{table:Eth-mchilowest-vs-temp} and Figure
\ref{fig:mchilowest-vs-temp_vesc-540}.}
\label{fig:rate_35degree_55degree_etaT100}
\end{figure}

\begin{figure}[thp]
\centering
\includegraphics[width=0.7\columnwidth]{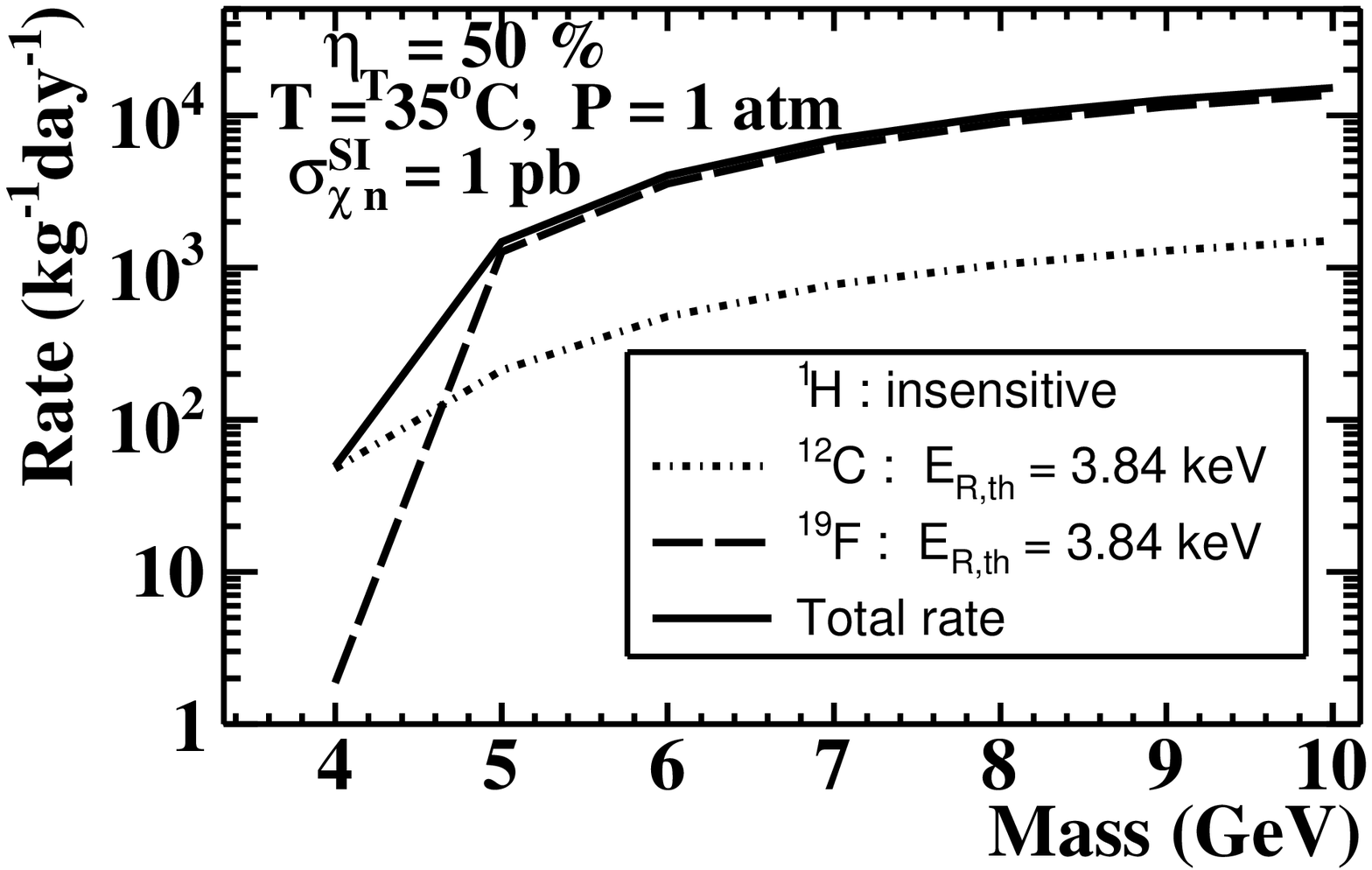}\\
\includegraphics[width=0.7\columnwidth]{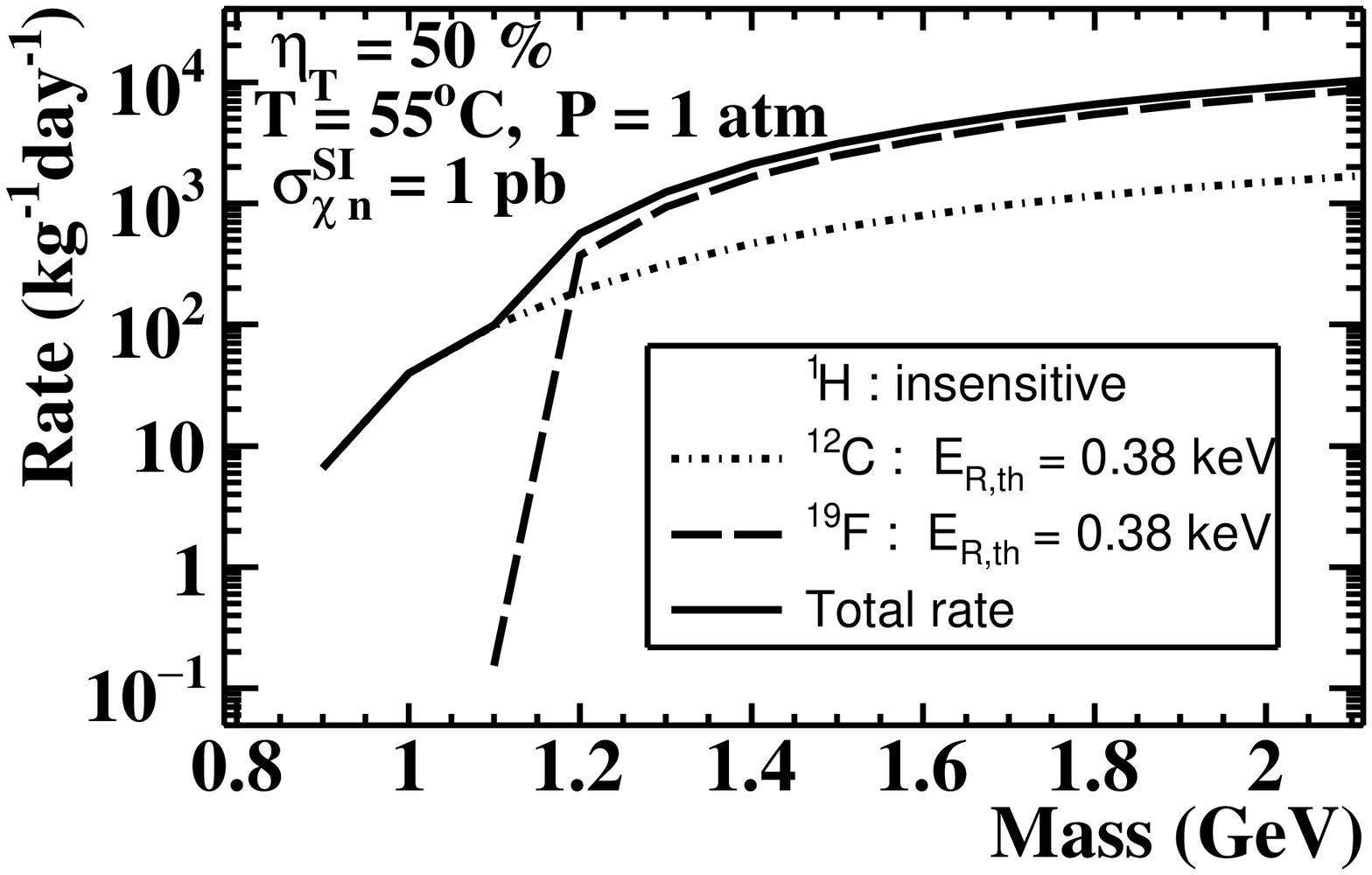}\\
\caption{Same as 
Figure \ref{fig:rate_35degree_55degree_etaT100} but for 
$\etaT=50\%$. Note that in this case there is no sensitivity to 
\hydrogen1 at both $35^\circ\,$C and $55^\circ\,$C --- the \hydrogen1 
sensitivity appears only at $T\protect\gsim 60^\circ\,$C in the case of 
$\etaT=50\%$; see Table \ref{table:Eth-mchilowest-vs-temp}.} 
\label{fig:rate_35degree_55degree_etaT50}
\end{figure}

The resulting rates (in units of kg$^{-1}$\ day$^{-1}$) as a function of 
WIMP mass for two different temperatures, $T=35^\circ\,$C and 
$T=55^\circ\,$C, for the case of $\etaT=100\%$, are shown in Figure 
\ref{fig:rate_35degree_55degree_etaT100} and those for the case of 
$\etaT=50\%$ are shown in Figure 
\ref{fig:rate_35degree_55degree_etaT50}.  

The event rates shown in these Figures scale 
linearly with the value of $\sigmachinSI$, and the number of events 
scale with the total exposure (kg.day). The WIMP-mass thresholds 
for contributions of different nuclei to the total event rates directly 
reflect the lowest WIMP-mass sensitivities shown in Table 
\ref{table:Eth-mchilowest-vs-temp} and Figure 
\ref{fig:mchilowest-vs-temp_vesc-540}. 

From Figures \ref{fig:rate_35degree_55degree_etaT100} and 
\ref{fig:rate_35degree_55degree_etaT50} we see that in the 
few GeV WIMP mass region the total event rates are generally dominated 
by contributions from recoiling \carbon12 and \fluorine19 nuclei. The 
sub-GeV WIMP mass region can be probed only by operating the detector at 
relatively high temperatures of $T\gsim 55^\circ\,$C. 
At such high temperatures, however, the detector will be 
sensitive to background gamma rays~\cite{c2h2f4-SunitaSahoo-2019}, which 
would give rise to bubble nucleation events through electron recoils 
(ER). This would seriously limit the potential of a \c2h2f4 detector 
for probing sub-GeV mass WIMPs unless effective means of rejection of 
the ER background can be found. 

In this context, recent work by the PICO 
collaboration~\cite{PICO-ER-recoil_2019} on electron recoils in 
\cthreef8 may point to a possible way forward. The   
analysis of gamma calibration data on \cthreef8 in 
Ref.~\cite{PICO-ER-recoil_2019} 
showed that bubble nucleation due to electron recoils in \cthreef8  
is better described in terms of a new nucleation mechanism that is 
driven by ionization through $\delta$-electron production rather than by 
localized energy deposition envisaged in the standard Seitz
heat spike model that well describes nucleation caused by nuclear
recoils. Consequently, bubble nucleation thresholds for electron 
and nuclear recoils scale differently with the 
thermodynamic operating conditions of the detector. In particular, the 
analysis of 
Ref.~\cite{PICO-ER-recoil_2019} suggests that it may be possible to 
minimize the nuclear recoil bubble nucleation threshold while maximizing 
the ER nucleation threshold by operating the detector at the lowest 
possible pressure. The operating temperature can then be tuned to the 
desired (low) nuclear recoil threshold, thereby making the detector 
largely insensitive to electron recoils without losing sensitivity to 
low energy nuclear recoil events. While it still remains to be 
demonstrated if this new model of ER bubble nucleation 
discussed in Ref.~\cite{PICO-ER-recoil_2019} for \cthreef8 would also be 
valid for other light element liquids, in particular \c2h2f4, the liquid 
of our interest in the present paper, below we estimate the level of 
sensitivity of a \c2h2f4 detector to low mass WIMPs under the assumption 
that appropriate thermodynamic operating conditions of the detector can 
be chosen so as to make such a detector insensitive to background gamma 
rays.    

\subsection{Sensitivity of \c2h2f4 detectors to low mass WIMPs under 
zero background assumption}
\label{subsec:sensitivity} 

Under the assumption of zero background, the standard Poissonian 90\% 
C.L.~upper limit on the WIMP-nucleon spin-independent cross section, 
$\sigmachinSIUL\,$, for zero observed number of events (which 
corresponds to a total of 2.3 expected number of events) is given by   

\begin{equation}
\frac{\sigmachinSIUL}{1\, {\rm pb}} \equiv \frac{2.3}{\mathcalR_{\rm 
exp}\,\mathcalE}\,,  
\label{eq:sigmachinSIUL_def}
\end{equation}
where $\mathcalE$ is the total exposure (in units of kg.day), and the 
expected total event rate $\mathcalR_{\rm exp}$ is calculated in units 
of (kg.day)$^{-1}$ for $\sigmachinSI = 1\pb$ (see  
Figures \ref{fig:rate_35degree_55degree_etaT100} and 
\ref{fig:rate_35degree_55degree_etaT50}). 

\begin{figure}[thp]
\centering
\includegraphics[width=0.7\columnwidth]{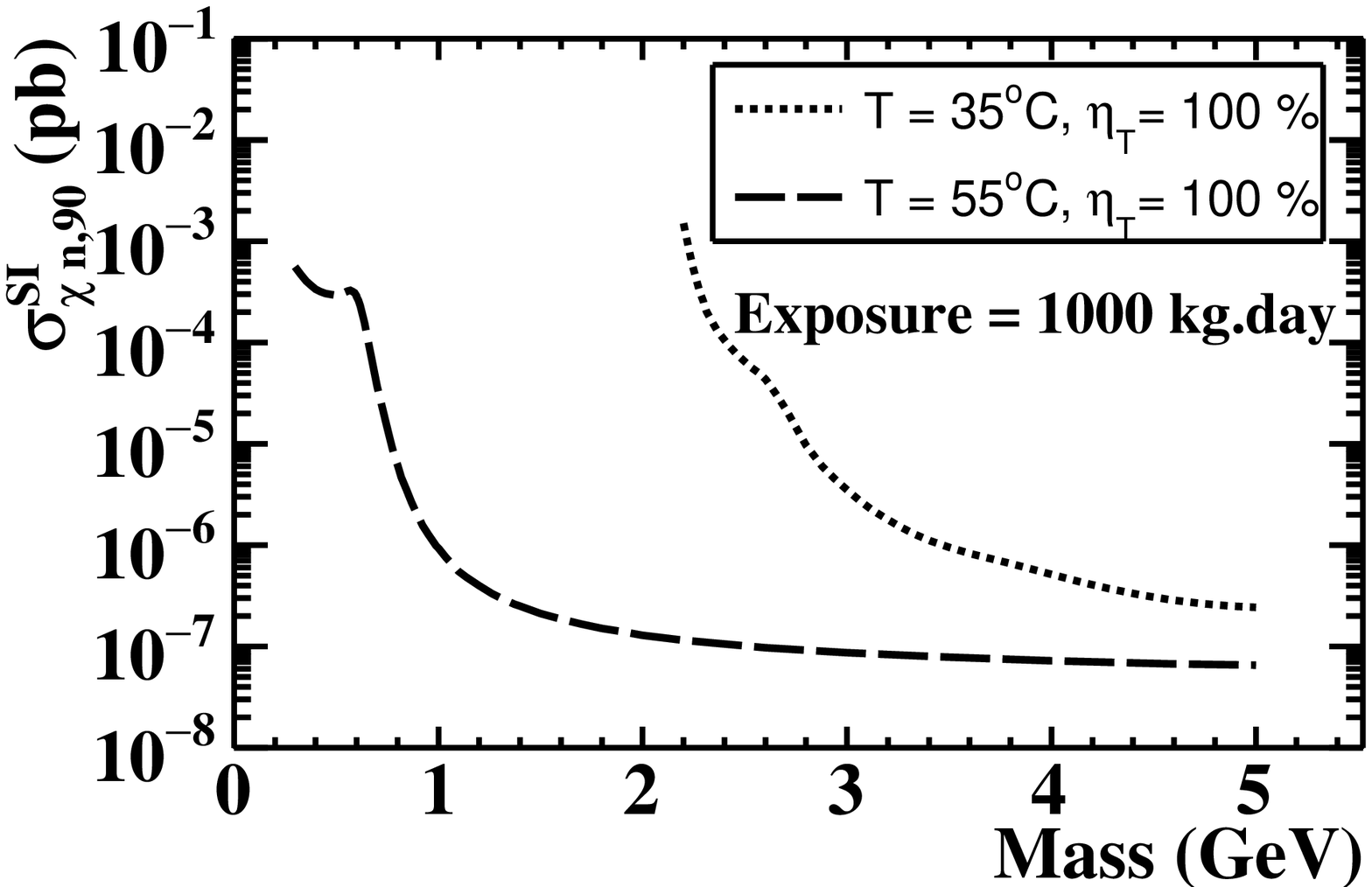}\\
\includegraphics[width=0.7\columnwidth]{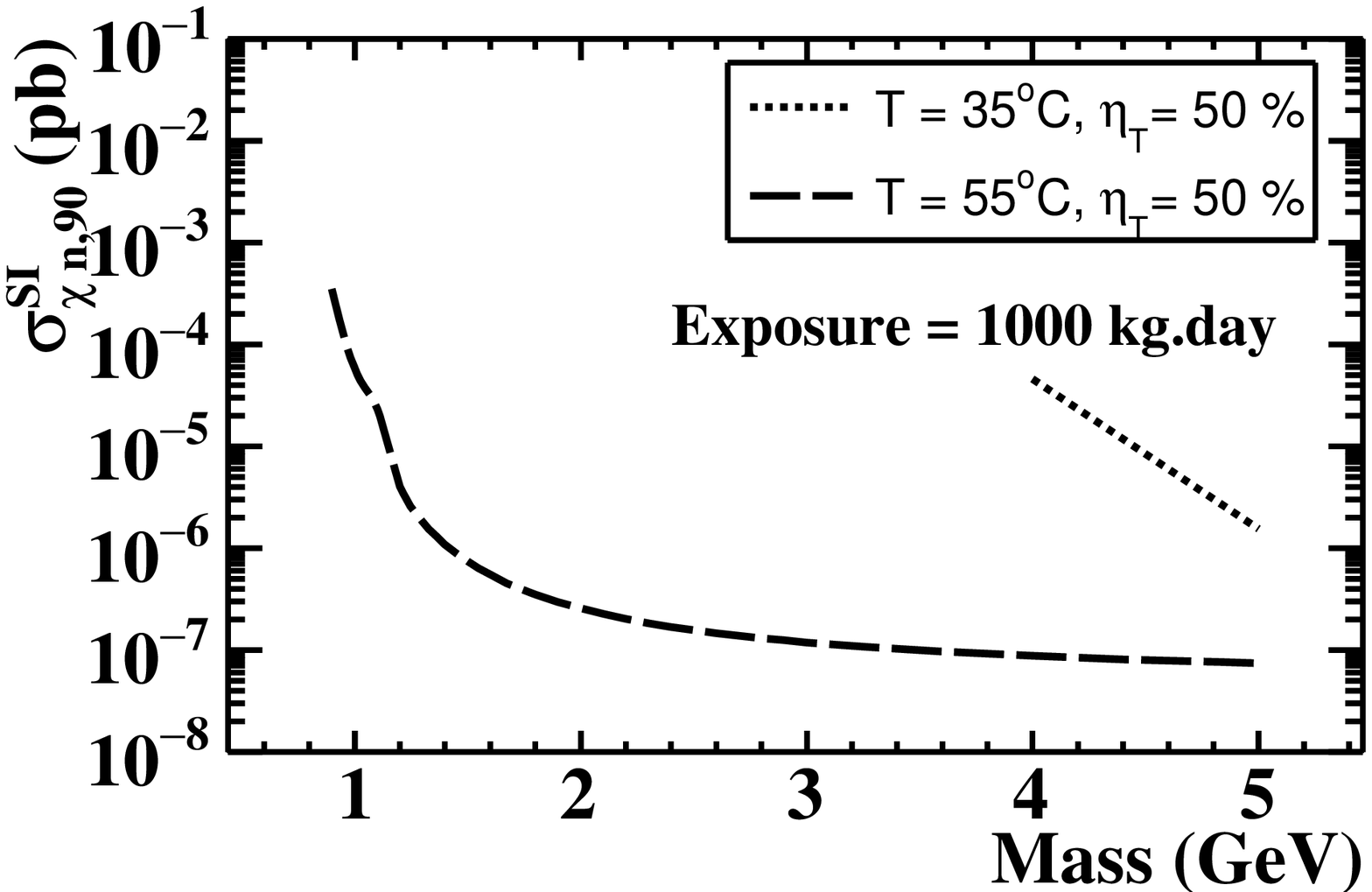}\\
\caption{90\% C.L. Poissonian upper limits on the spin-independent 
WIMP-nucleon cross section, $\sigmachinSIUL$, as a function of WIMP mass 
for zero observed events under the assumption of zero background for a 
total exposure of 1000 kg.day, for operating temperatures 
$T=35^\circ\,$C and 
$55^\circ\,$C, and $\etaT=100\%$ (upper panel) and $\etaT=50\%$ (lower 
panel).}
\label{fig:exclusion_plot_etaT100_etaT50} 
\end{figure}

Figure \ref{fig:exclusion_plot_etaT100_etaT50} shows $\sigmachinSIUL$ as 
a function of WIMP mass (sub-GeV -- few GeV) for a total exposure of 
$\mathcalE=10^3\,$ kg.day for $\etaT=100\%$ and 
$\etaT=50\%$, in each case for two values of the 
operating temperature, $T=35^\circ\,$C and $55^\circ\,$C. 

It is seen that, in the situation of zero background, a \c2h2f4 SLD with 
$\etaT=100\%$ operated at $55^\circ\,$C (corresponding to a Seitz 
threshold of 0.19 keV) with a total exposure of 1000 kg.day, for 
example, would be able to probe WIMPs of masses 5, 3, 2 and 1 GeV at the 
sensitivity levels of $\sigmachinSIUL = 
6.5\times 10^{-8}$, $8.7\times10^{-8}$, $1.3\times10^{-7}$ and 
$9.3\times10^{-7}\pb$, respectively. Note that in the few-GeV mass 
region the event rates at $55^\circ\,$C are dominated by \fluorine19 
recoils, with \carbon12 recoils making sub-dominant and \hydrogen1 
recoils making negligible contributions. In the sub-GeV mass region, at 
$55^\circ\,$C, the sensitivities are at the levels of 
$6.2\times10^{-6}$, 
$3.7\times10^{-5}$, $2.9\times10^{-4}$, and $5.6\times10^{-4}\pb$ at 
$\mchi=0.8\,, 0.7\,, 0.5\,$ and 0.3 GeV, respectively. Note that, 
at $55^\circ\,$C, the \carbon12 and \fluorine19 recoils are unable to 
cause bubble nucleation events for WIMPs of masses $\lsim 0.6\GeV$, 
and the limits on $\sigmachinSI$ for these WIMP masses come 
from WIMPs scattering on \hydrogen1 only. 

At the lower temperature of $35^\circ\,$C, again for $\etaT=100\%$ 
(corresponding to a Seitz threshold of 1.92 keV), 
the sensitivities in the few GeV WIMP mass region worsen in comparison to 
those at $55^\circ\,$C, with $\sigmachinSIUL = 2.5\times 10^{-7}$, 
$5.2\times10^{-7}$, $3.5\times10^{-6}$  and $2.2\times10^{-5}\pb$
at $\mchi=5\,, 4\,,$ 3 and 2.7 GeV, respectively.  
Again, these limits come mainly from \fluorine19 recoils, with \carbon12 
recoils making sub-dominant and \hydrogen1 recoils making no 
contribution. In the WIMP mass region $2.2 \lsim \mchi \lsim 2.6$ GeV, 
the \carbon12 recoils make the dominant contribution to the event rate, 
giving a sensitivity at the level of $\sim 1.5\times 10^{-3}\pb$ at 
$\mchi\sim 2.2$ GeV, the lowest WIMP mass that can be probed at 
$35^\circ\,$C in the case of $\etaT=100\%$. 

In the case of $\etaT=50\%$, at $T=35^\circ\,$C, in the few GeV 
WIMP-mass region, the $\sigmachinSIUL$ sensitivities are at the levels 
of 
$4.6\times10^{-5}\pb$ and $1.6\times10^{-6}\pb$ at $\mchi=4$ and 5 GeV, 
respectively. There is no sensitivity to sub-GeV WIMPs at this 
temperature. At the higher temperature of $T=55^\circ\,$C, the 
sensitivities improve with reduction of thresholds, with values of 
$7.5\times10^{-8}$, $2.6\times10^{-7}$ and $5.8\times10^{-5}\pb$ 
at $\mchi=5$, 2 and 1 GeV, respectively, for example, with dominant 
contributions to the event rates coming from \fluorine19 recoils for  
$\mchi=5$ and 2 GeV, and from \carbon12 recoils for $\mchi=1$ 
GeV. There is only marginal sensitivity to sub-GeV mass WIMPs, which can 
be reached only at $T\gsim 60^\circ\,$C in this case of $\etaT=50\%$. 

\section{Summary and Conclusions}\label{sec:summary}
In this paper we have theoretically studied the potential of 
a superheated liquid detector (SLD) with a hydrogen containing liquid, 
namely, tetrafluoroethane (C$_{2}$H$_{2}$F$_{4}$) (b.p.
$-26.3^\circ\,$C), as the active target material for probing low 
(sub-GeV -- few GeV) mass WIMP candidates of dark matter. In a \c2h2f4 
SLD the recoiling \hydrogen1, \carbon12 and \fluorine19 nuclei arising 
from elastic scattering of the Galactic WIMPs on these nuclei  
can give rise to detectable bubble nucleation events if the recoil 
energies are above certain threshold energies. The latter  
are determined by the ``Seitz condition" that a fraction 
$\etaT<1$ (called the ``thermodynamic efficiency") of the energy 
deposited by a recoiling nucleus in the liquid over a length scale of 
$2\Rc$ (where $\Rc$ is a critical length) has to be greater 
than or equal to a certain critical energy $\Ec$. Both $\Rc$ and $\Ec$ 
are characteristics of the liquid under consideration, and depend on the 
operating temperature and pressure of the superheated liquid, thus 
making the threshold energies dependent on temperature at a given 
operating pressure. 

We have discussed the calculation of the bubble nucleation threshold 
energies of the WIMP-induced recoiling \hydrogen1, \carbon12 and 
\fluorine19 nuclei in \c2h2f4 using the Seitz condition for various 
values of the parameter $\etaT$. Guided by the results from 
the PICO experiment~\cite{pico-60-C3F8-PRD2019} on  
the experimental ranges of possible values for the bubble nucleation 
threshold energies and the nucleation efficiencies above their 
respective thresholds for \carbon12 and 
\fluorine19 recoils in \cthreef8, we have shown the results 
for the event rates and WIMP-mass sensitivities for \c2h2f4 for two 
possible representative different values of $\etaT$, namely, 
50\% and 100\%, for illustration. For \hydrogen1 recoils in \c2h2f4, in 
absence of any direct experimental results on their nucleation 
efficiency curves, we have assumed 100\% bubble 
nucleation efficiency at threshold as in previous 
studies~\cite{PICO-thesis-Tardif-2018}. 

In general, the bubble nucleation 
threshold energies of all nuclei decrease with increasing temperature 
irrespective of the value of $\etaT$. In the ideal case of 
$\etaT=1$, by operating the detector at a relatively high temperature of 
$T=60^\circ\,$C, for example, nucleation thresholds of $\sim$ 0.1  
keV (same for \hydrogen1, \carbon12 and \fluorine19) can be reached,  
allowing sub-GeV WIMP masses down to $\sim$ 140, 430 and 540 MeV (due to 
\hydrogen1, \carbon12 and \fluorine19 recoils, respectively) to be 
probed. At such high temperatures, however, the detector will be 
sensitive to background gamma rays~\cite{c2h2f4-SunitaSahoo-2019}. At a 
lower temperature of $35^\circ\,$C, for example, where the detector is 
expected to be insensitive to gamma rays, \carbon12 and \fluorine19 
thresholds of $\sim$ 2 keV would be possible, making the detector 
sensitive to WIMPs of masses down to $\sim$ 2.2 GeV. However, 
sensitivity to \hydrogen1 recoils would be lost at this temperature. 

In a perhaps more realistic situation of $\etaT=0.5$, for example, 
\carbon12 and \fluorine19 thresholds move up, for reasons explained in 
section \ref{subsec:results_Eths}, by a factor of 2 compared 
to their values for the case of $\etaT=1$ at all temperatures of 
interest considered here, with lowest sensitive WIMP-mass moving 
upwards by factors between $\sim$ 1.4 and 1.5 compared to the case of 
$\etaT=1$. Sensitivity to \hydrogen1 recoils now appear only at 
temperatures $T\gsim 60^\circ\,$C\  where, again, the 
detector will be sensitive to background gamma rays.    

From the above discussions we see that few-keV level recoil energy 
thresholds possible for bubble nucleation in \c2h2f4 SLDs by recoiling 
\carbon12 and \fluorine19 nuclei in \c2h2f4 at gamma-ray insensitive 
temperature regions of $T\lsim 35^\circ\,$C, for example, have the 
potential to allow WIMPs in the few-GeV mass range to be probed at a 
WIMP-nucleon spin-independent cross section sensitivity (90\% C.L.) 
levels better than $\sim 4.6\times 10^{-5}\pb$ at WIMP masses down to 
$\sim$ 4 GeV with a total exposure of $\sim$ 1000 kg.day, 
provided that thermodynamic efficiency $\etaT$ that determines the 
bubble nucleation thresholds for the recoiling \hydrogen1, \carbon12 
and \fluorine19 nuclei in \c2h2f4 is $\sim$ 50\% or higher. On the other 
hand, sensitivity 
to sub-GeV WIMP masses generally requires the detector to be 
sensitive to the WIMP-induced \hydrogen1 recoils, which in turn requires 
the detector to be operated at gamma sensitive temperatures  
$T\gsim50^\circ\,$C and $\etaT$ substantially larger than 50\%. 

As discussed above, the sensitivity of a \c2h2f4 detector to background 
gamma rays at high temperatures (corresponding to low Seitz thresholds),  
from sources both external as well as intrinsic to the detector 
material, poses a serious challenge for probing sub-geV WIMP masses 
unless effective means of making the detector insensitive to background 
gamma rays, which would cause bubble nucleation due to electron recoils, 
can be found. In this regard, a possible way forward may come from a 
recent work by the PICO collaboration~\cite{PICO-ER-recoil_2019} which, 
based on gamma calibration data for \cthreef8,  
suggests a different mechanism of electron recoil induced bubble 
nucleation in \cthreef8 than that envisaged in the standard 
Seitz heat spike scenario of bubble nucleation due to 
nuclear recoils. If this new scenario of gamma induced 
bubble nucleation in \cthreef8 can be demonstrated to be valid for other 
light element liquids, in particular \c2h2f4, then following the 
prescription in Ref.~\cite{PICO-ER-recoil_2019} it may be 
possible to push the electron recoil bubble nucleation thresholds to 
relatively higher energies while simultaneously lowering the 
nuclear recoil bubble nucleation thresholds to relatively lower energies 
by appropriately choosing the operating pressure and temperature of the 
detector, thereby making the
detector largely insensitive to ER events without losing sensitivity to
low energy hydrogen recoil events. We wish to explore this possibility 
in a future work. 

\vspace{0.5cm}
\noindent {\bf Acknowledgments :} 
We thank members of the PICO collaboration, in particular, Alan 
Robinson, Ubi Wichoski and Viktor Zacek, for useful comments. We wish 
to thank an anonymous referee for making valuable comments 
on a previous version of the manuscript, which has resulted in major 
changes and what we believe is a substantially improved version  
of the manuscript. This work is supported by grants under the CAPP-II 
project funded by the Dept.~of Atomic Energy (DAE), Govt.~of India. One 
of us (PB) acknowledges support under a DAE Raja Ramanna Fellowship.

\end{document}